\documentclass[fleqn,usenatbib]{mnras}

\usepackage{newtxtext,newtxmath}

\usepackage[T1]{fontenc}

\DeclareRobustCommand{\VAN}[3]{#2}
\let\VANthebibliography\thebibliography
\def\thebibliography{\DeclareRobustCommand{\VAN}[3]{##3}\VANthebibliography}

\usepackage{graphicx}	
\usepackage{amsmath}	
\usepackage{arydshln} 
\usepackage{threeparttable} 
\usepackage{booktabs}
\usepackage{xcolor}

\newcommand{\CenterCell}[1]{\begin{tabular}{@{}c@{}}#1\end{tabular}}
\newcommand{\barolo}{$^{\text{3D}}$\textsc{Barolo}}
\newcommand{\galpak}{GalPaK$^{\text{3D}}$}
\newcommand{\qubefit}{\texttt{qubefit}}
\newcommand{\cii}{[C\,{\sc ii}]}

\title[High-z galaxy kinematics 3D tools' biases]{Kinematics of synthetically observed high-$z$ rotating disks: reliability and biases of 3D fitting tools}

\author[M. Yttergren et al.]{
    M. Yttergren,$^{1}$\thanks{E-mail: madeleine.yttergren@chalmers.se}
    K. K. Knudsen,$^{1}$
    J. Molina,$^{2,3}$
    G. C. Jones,$^{4,5}$
    K. Kade,$^{1}$
    J. Scholtz,$^{4,5}$
    A. Bewketu Belete$^{1}$
\\
    $^{1}$Department of Space, Earth and Environment, Chalmers University of Technology, SE-412 96 Gothenburg, Sweden\\
    $^{2}$Instituto de F\'isica y Astronom\'ia, Universidad de Valpara\'iso, Avda. Gran Breta\~na 1111, Valpara\'iso, Chile\\
    $^{3}$Millennium Nucleus for Galaxies (MINGAL)\\
    $^{4}$Kavli Institute for Cosmology, University of Cambridge, Madingley Road, Cambridge CB3 0HA, UK\\
    $^{5}$Cavendish Laboratory, University of Cambridge, 19 JJ Thomson Avenue, Cambridge CB3 0HE, UK
}

\date{Accepted 2025 August 29. Received 2025 August 07; in original form 2025
April 10}

\pubyear{\the\year{}}

\begin{document}
\label{firstpage}
\pagerange{\pageref{firstpage}--\pageref{lastpage}}
\maketitle

\begin{abstract}
    Resolved high-redshift galaxy gas kinematics is a rapidly evolving field driven by increasingly powerful instrumentation. 
    However, the resolution and sensitivity still impose constraints on interpretation. 
    We investigate the uncertainties inherent to high-$z$ galaxy kinematical analysis by modelling a suite of rotating disk galaxies, generating synthetic interferometric ALMA observations, and fitting them with the 3D-kinematical tools \barolo, \galpak, and \qubefit. 
    We present the recovered 3D-fitted kinematical parameters to assess their reliability, quantify the range of values possible for individual source studies, and establish the systematic biases present for observed samples. 
    The $V/\sigma_{\rm V}$ ratio, which indicates how dynamically cold a system is, is of particular importance and depends on the choice of 3D-fitting tool. 
    On average, \barolo\, and \qubefit\, slightly overestimates $V/\sigma_{\rm V}$ ($<1\sigma$) and \galpak\, underestimates it ($<2\sigma$). 
    Therefore, all three tools are reliable for kinematical studies of averages of high-redshift galaxy samples. 
    The value range possible for individual sources is significant, however, even more so for samples of not purely rotation dominated sources.
    To determine whether an observed galaxy is rotation dominated enough to be fitted with a 3D-kinematical tool, $V/\sigma_{\rm V}$ can be extracted directly from the observed data cube, with some caveats. 
    We recommend that the median offsets, value ranges, and tool-dependent biases presented in this paper are taken into account when interpreting 3D-fitted kinematics of observed high-redshift galaxies.  
\end{abstract}




\section{Introduction}
The widespread presence of disk-like galaxies at $z \approx 1-3$ is considered a key probe of internal mechanisms regulating the life cycle of star-forming galaxies and their evolution across cosmic time \citep{forsterschreiber2020}. About half of the current stellar mass observed in galaxies today was formed during an epoch when these systems appeared clumpy and irregular in the restframe UV and optical \citep{swinbank2012,conselice2014}, but their kinematics and global light distribution were largely regular \citep[e.g.][]{wuyts2011b,vanderwel2014,wisnioski2019}.
Smooth modes of gas accretion, star formation, and outflows set conditions for equilibrium growth that controlled the stellar mass build-up of galaxies \citep[e.g.][]{genzel2006,dekel2009}.

Surveys spatially resolving galaxy spectra using integral field units (IFUs) have been pivotal in establishing this picture \citep[e.g.][]{forster2009, swinbank2012,wisnioski2015,stott2016,turner2017,forster2018}, and complementary interferometric observations unveiling the cold molecular interstellar medium (ISM) gas phase at high redshift have provided further supporting evidence \citep[e.g.][]{genzel2013,molina2019,ubler2018,genzel2023,rizzo2023}. Surprisingly, recent reports suggest a high prevalence of dynamically cold rotating disks at even higher redshifts, where the dynamical ''coldness'' of a system is derived from the $V_{\rm rot}/\sigma_{\rm V}$ ratio of gas tracers such as H$\alpha$ and \cii \citep{rizzo2020,rizzo2021,pope2023,kohandel2024,rowland2024,scholtz2025}. These observations have been further corroborated by {\it James Webb Space Telescope} ({\it JWST}) imaging analysis \citep{robertson2023}, suggesting that galaxies were building their stellar mass in a quasi-equilibrium state \citep[e.g.][]{krumholtz2010} even at earlier epochs. 
However, {\it JWST} observations also show the presence of companion galaxies and pre-coalescence mergers \citep[e.g.][]{jones2024,scholtz2024,lamperti2024}, raising doubts about the accuracy of the tools currently used for identifying and characterizing galaxy disks from three-dimensional datasets. 
For example, \citet{simons2019} concluded from synthetic IFU observations at $z\approx2$ that for close-pair mergers the risk of classifying a merger as a disk can be as high as 100\% depending on the specific disk classification criterion adopted (and weakly dependent on the separation of the merging galaxies). \citet{rizzo2022} showed that distinguishing between a rotating disk and a merger, and correctly classifying them with the current data quality and methods, is impossible for all but the highest data quality, achievable only for a small sample of the brightest galaxies. 
Unfortunately, the limited sensitivity and resolution of current observations present challenges in characterizing the morphology and kinematics of the ISM in high-redshift galaxies.

The current tools used to classify and characterize the gas kinematics of high-redshift galaxies are far from flawless. 
These tools face a difficult task: modelling galaxy data with the relatively large point spread function (PSF) or synthesised beam size, which smooths out and circularizes any galaxy structure. 
The observed rotation velocity fields also appear more regular than they actually are, and the velocity dispersion is artificially increased due to the broadening of the emission lines by unresolved velocity gradients, plus the finite spectral resolution of the instrument (see \citealt{glazebrook2013}, for a review). 
In some cases, due to the smoothing, even merging systems can be classified as disk-like galaxies \citep{smit2018, simons2019, rizzo2022}. 
The tools developed for analysing the ''2D'' kinematic fields correct for ''beam-smearing'' and instrument spectral broadening effects \citep[e.g.][]{swinbank2012,wisnioski2015,stott2016,levy2018}. 
However, since the convolution with the PSF or synthesized beam is flux-weighted, it affects data cubes as a whole, implying that a more appropriate solution to account for these effects is to model the datasets in three dimensions. 

Over the last decade, several ``3D'' galaxy kinematics modelling tools have been developed to overcome these challenges and fully exploit the data cubes delivered by IFUs and radio interferometers (e.g., the Atacama Large Millimetre/sub-millimetre Array--ALMA). Many of these ``3D-kinematical tools'' are publicly available, and a non-exhaustive list includes  {\tt TiRiFiC} \citep{jozsa2007}, {\tt KinMS} \citep{davis2013, davis2017}, \barolo\, \citep{diteodoro2015,diteodoro2021}, \galpak\, \citep{galpakcode2015,bouche2015}, \qubefit\,  \citep{neeleman2020thecode,neeleman2021}, and DysmalPy \citep{davis2011,davis2013,price2021,lee2024}. The major difference among these tools lies in the adopted approach to model the data, which can be largely arranged into routines that use multiple tilted rings ({\tt TiRiFiC} and \barolo) or parametrically defined profiles (\galpak, {\tt KinMS}, \qubefit\, and DysmalPy) to fit the data. Parametric models assume how the parameters--such as the intensity, rotation velocity, and disk thickness/dispersion profiles--vary with radius before fitting, unlike the multiple tilted-ring models. However, all of these tools assume that galaxies are well-described by axisymmetric geometries, which may very well not be true for high-redshift systems. The blurring or beam-smearing of the data limits any detailed decomposition of the systems into their constituents, meaning the characterization of the substructures of the high-redshift ISM remains nearly uncharted, with a few exceptions made possible by strong gravitational lensing \citep{motta2018,dessauges-zavadsky2019,kade2024}. 

\citet{lee2024} benchmark DysmalPy, \barolo, and \galpak\, against each other using samples of disks, created through each of the respective tools, DysmalPy, \barolo, or \galpak, and modelled after main-sequence star-forming galaxies at $z=1-3$. 
They find a clear correlation between the accuracy of the recovered rotation velocity and velocity dispersion and the method of creation of the input galaxies. The intrinsic template mismatch can result in up to a factor 2 offset of the recovered kinematical parameters. 
Furthermore, \citet{lee2024} find that all three tools recover the rotation velocity accurately but find a substantial scatter in the velocity dispersion. 
For \barolo\, they note in particular a risk of over-masking, especially at low velocity dispersions, resulting in underestimation of the line widths. 

This paper aims to comprehensively study the capabilities of 3D-kinematic tools in recovering the kinematic properties of the gas in high-redshift galaxies. 
We focus on building mock galaxy data from physical thin disk models and synthetic interferometric ALMA observations of these systems within a common framework. Our study focuses on rotating disks with multiple input geometries and data quality setups, which we fit using the public 3D-kinematic tools: \barolo, \galpak, and \qubefit. 
These three tools were selected, partly to constrain the parameter space, and primarily due to their respective claims of being able to perform well on poorly resolved data, i.e., high-redshift galaxy data. 
Our main goal is to gain clear insights into what aspects of the kinematics can be trusted most, which will allow us to better understand the interpretation of transitional qualities and processes present in high-redshift galaxies. 
We do this by providing a range of percentage offsets on the kinematics expected when 3D-fitting observed galaxies, and thereby allowing observers to avoid over-interpreting their results. 
This paper begins with a description of the rotating galaxy disk models in section~\ref{ch:sim}, the simulation process of synthetic interferometric ALMA observations in Section \ref{ch:simalma} and the 3D-kinematical tool's setup in Section \ref{ch:3Dfitting}. 
Section \ref{ch:results} details the kinematical parameters obtained via fitting the synthetically observed galaxy disks using \barolo, \galpak, and \qubefit. 
In Section \ref{ch:further_discussion}, we discuss the considerations required when using the 3D-kinematical tools, and the range of recovered kinematical parameters focusing on $V/\sigma_{\rm V}$, its method of derivation and impact on the conclusion of dynamically cold disks. We conclude with a summary in Section \ref{ch:sum}. 
Throughout this work, we assume a $\Lambda$CDM cosmology with $\Omega_{\rm m} = 0.3$, $\Omega_\Lambda = 0.7$, and $H_0 = 68\,$km\,s$^{-1}$\,Mpc$^{-1}$ \citep{planck15cosmology}.

\section{Disk modelling}
\label{ch:sim}  
We adopt simple analytic models of thin disk galaxies to provide a controlled environment for testing and separating the effects of the different influential components. 
Assuming a thin disk is an ideal case that is commonly adopted in kinematical analyses, including in the defaults of the kinematical tools investigated in this paper. 
It is a sufficient assumption here as we fit the rotation velocity and dispersion independently and do not predict the galaxy height (while a thick disk correlate the height and radius of the galaxy to the observed kinematics ($h/R \sim \sigma/V$)). 
The models differ in which parametric profiles are used to describe the radial dependence of the brightness, rotation velocity, and velocity dispersion along the major axis of the disk. 

Note that our models rely on theoretical profiles detailing physical rotating disks derived from two different gravitational potentials. 
We begin this section by presenting the two different theoretical disk models, thereafter we delve into the modelling framework and how we incorporate velocity dispersion.

\subsection{Matter distribution and rotation velocity} \label{ch:theoreticalgalaxyprofiles}
There are multiple different theoretical galaxy disks in literature \citep[see for example][]{freeman1970,binneytremaine}. 
For our models we choose two different disks defined by two separate gravitational potentials. 
From the gravitational potentials the flux and the rotation velocities of the respective disks follow as a function of radius.

\subsubsection{Exponential disk}\label{ch:exp_disk}
One of the most well known, and often observed, brightness distributions in galaxy disks is the exponential (``exp''), which has a brightness profile that peaks in the central region and then decreases exponentially. 
The exponential disk is described by: 
\begin{equation}\label{eq:exp_sersic}
    I(r) = I_0 \exp (-r/l),
\end{equation}
where $I_0$ is the brightness of the central pixel, and $l$ is the scale length of the disk (current estimates of the Milky Way's scale length is 2-3 kpc) \citep{sersic1963,freeman1970,binneytremaine}. 

The gravitational potential of such a system is, in the galactic plane: 
\begin{equation}
\begin{split}
    \Phi (r,0) & = -4\pi G \Sigma_0 \int^R_0 da \frac{aK_1(a/l)}{\sqrt{r^2 - a^2}} \\ & = -\pi G \Sigma_0 r [I_0(x)K_1(x) - I_1(x)K_0(x)],
\end{split}
\end{equation}
where $x=r/(2l)$, $\Sigma_{0}$ is the surface mass density, $I_{n}$, $K_{n}$ are the modified Bessel functions of n-th order, and $l$ is the scale length. \citep{binneytremaine}. 
The corresponding rotation curve to such a gravitational potential rises until 2.15$l$, after which it declines. Note that in this paper we refer to this profile as ``rising to declining'' (``risDec''). 
The rotation curve is 
\begin{equation}\label{eq:risingtodeclining}
    V(r) = \sqrt{\frac{4 \pi G \Sigma_{0}}{l}} l x^2 \left[ I_0(x)K_0(x) - I_1(x)K_1(x) \right] ^{1/2}.
\end{equation}
In our models we set the surface mass density to $\Sigma_{0}=4125\,\text{M}_{\odot}\text{pc}^{-2}$ as this corresponds to a $V_{\rm max}$ of 250 km/s, which matches the rotation velocity values chosen for the other investigated rotation curves.

\noindent
\subsubsection{Cored logarithmic disk} \label{ch:corelog_disk}
The other gravitational potential we adopt is the logarithmic potential, which we here often refer to as the cored logarithmic (``corelog'') disk. 
This type of gravitational potential is of particular interest as, while it does not require modelling of both a baryonic disk and a dark matter halo, the shape of the gravitational potential results in a baryonic disk that behaves very similar to such a dual system's disk. 
The logarithmic potential corresponds to a disk with slightly increased central mass, a core, with a $r^{-2}$ decrease thereafter, and the rotation curve is rising at lower radii and then flattens out -- we call this rotation curve ``rising to flat'' (``risFlat''). 
This kind of rotation curve better matches observations, and it was the observations of rising-to-flat rotation curves that was one of the first pieces of evidence for dark matter in galaxies \citep{rubin1970}. 

The logarithmic gravitational potential is \citep{binneytremaine}): 
\begin{equation}
    \Phi(r) = V_{\rm max}^2 \frac{\ln[r_{\rm d}^2 + r^2]}{2}, 
\end{equation}
where $r_{\rm d}$ is the kinematical scale radius. 
The brightness profile of this system is:
\begin{equation}
    I(r) = I_0 \frac{r_{\rm d}^4}{(r_{\rm d}^2 + r^2)^2}, 
\end{equation}
where $I_0$ is the brightness of the central pixel. 
And the rotation curve is described by: 
\begin{equation}\label{eq:risingtoflat}
    V(r) = \frac{V_{\rm max} r}{\sqrt{r_{\rm d}^2 + r^2}},
\end{equation}

The kinematical scale radius, $r_{\rm d}$, relates to the total gravitational potential of the system, which means that in our case this value is comparable to the disk scale length, $l$, used in the exponential disk. Eq.~\ref{eq:exp_sersic}. 
Therefore our models are constrained to $r_{\rm d}=l$.

\subsection{Velocity dispersion} \label{ch:modelling_velocity_dispersion}
Infinitesimally thin modelled disks have no physical velocity dispersion inherent in the disks themselves apart from any assumption we make regarding the gas' inherent turbulent motions due to Brownian motion and temperature (there is also a contribution from the change in line-of-sight velocity across the individual pixel/spaxel when the disks are observed); however, this is not a fully accurate description of galaxies. 
Observations have found values for star-forming galaxies at $z\approx1-4$ in the range of $50-100$ km/s with a $V_{\rm max}/\sigma_{\rm V}$ of 1--10 \citep[e.g.][]{law2009, genzel2011, fs2018, johson2018, wisnioski2019, pillepich2019, kohandel2020, herreracamus2022, birkin2024, romanoliveira2023, pope2023}. 
In our standard setup, we have therefore chosen $\sigma_{\rm V}=50$ km/s and $V_{\rm max}=250$\,km/s, resulting in $V_{\rm max}/\sigma_{\rm V} = 5$. 
And we further vary the velocities to achieve a range of $V_{\rm max}/\sigma_{\rm V} = 0.5 - 10$. 
We choose two different velocity dispersion profiles as a function of radii, one constant, to mimic an ideal disk with a homogeneously distributed gas at a constant temperature, and an exponential profile, as gas dispersion has in some cases been observed to increase towards the centre of galaxies \citep[e.g.][]{rizzo2022,rizzo2023,romanoliveira2023,birkin2024}. 
The dispersion profile is therefore artificially modelled as either constant, $\sigma_{\rm V}(r) = \sigma_{0}$, or 
$\sigma_{\rm V}(r) = \sigma_{0} \exp (-r/r_{\sigma})$, 
where we in our case set the characteristic radius, $r_{\sigma}$, equal to the scale radius, $r_{\sigma} = r_{\rm d}$, and $\sigma_{0}$ so that $\sigma_{\rm V}(r_{\rm d}) = \sigma_{0} = 50$\,km/s, to limit the parameter space.

\subsection{The modelling framework} \label{ch:modelling_framework}
The modelled galaxies are infinitesimally thin axisymmetric rotating disks whose matter distribution, rotation velocity and velocity dispersion follow the radial formulae presented in Section \ref{ch:theoreticalgalaxyprofiles} and \ref{ch:modelling_velocity_dispersion}. 

The modelling process is as follows:
\begin{enumerate}
    \item Define a three dimensional normal vector for plane that the modelled disk resides in. 
    \item Define a x--y pixel grid with a pixel size, onto which the disk is projected while storing the xyz values of each projected point in units of arcseconds. 
    \item Calculate the flux, line-of-sight velocity, and velocity dispersion in each pixel based on the choice of brightness, rotation curve, and dispersion profiles. Note that apart from user defined galaxy parameters, these functions/profiles depend only on the distance from the centre of the disk to the pixel in question. 
    The line-of-sight velocity, $V_{\rm los}$, is derived from the rotation speed, $V(r)$ as:
\begin{equation}\label{eq:Vlos_to_Vrot}
    V_\mathrm{los} = V(r)\sin (i) \cos({\rm PA}),
\end{equation}
where $i$ is the inclination and PA is the position angle. 
\begin{itemize}
    \item As each of these three parameters vary as a function of radii, the change across a single pixel can be large, depending on the pixel size. To account for this, we create a 10$\times$10 sub-pixel grid and set each parameter to be the average of multiple computed values within one pixel. 
    \item Furthermore, the change in line-of-sight velocity across the pixel broadens the emission line, i.e., causes an increase in the observed velocity dispersion. This is accounted for by quadratically adding the standard deviation of the 100 computations of the line-of-sight velocity in each pixel to the respective velocity dispersion.
    \item Throughout this process, flux is conserved. 

\end{itemize}

    \item
    Create the galaxy cube by constructing an artificial Gaussian emission line from the flux, line-of-sight velocity, and velocity dispersion values of each pixel. The modelled galaxy disk is saved as a classical three dimensional Flexible Image Transport System (FITS) file with two spatial dimensions and one spectral.

\end{enumerate}

The x--y plane is the plane of the sky, the z-axis is the line-of-sight. 
The inclination of the galaxy is the angle between the line-of-sight and the normal of the plane of the galaxy, so that a 0$^{\text{o}}$ inclination is a face-on galaxy and a 90$^{\text{o}}$ inclination is an edge-on galaxy. 
The position angle is the angle between the y-axis and the major axis of the galaxy, positive values counter-clockwise. 

A large sample of disk models with different combinations of profiles and settings were created, from this we chose a suitable smaller sample of disks to cover a range of behaviours. 
The final sample can be split into two main investigation topics, the first half focuses on different profile combinations and the second half focuses on the impact of rotation versus dispersion-dominated systems. 
The settings of these disks are listed in Table \ref{table:simgals_for_simalma}.

\subsection{Connection to observational values}
Our disk models are of \cii\ 1900.536 GHz, as \cii is a commonly used fine-structure emission line for resolved kinematics studies of high-$z$ galaxies \citep[see e.g.][]{kohandel2019,rizzo2020,jones2021,kade2024,devereaux2024,telikova2024}. 
\cii\ is expected to trace warm and cold gas. 
We adopt a redshift of 5.1, which also sets the physical size our galaxy inhibits per arcsecond, 6.365 kpc~arcsec$^{-1}$. 
Assuming an integrated \cii\ line luminosity over the galaxy of $5\cdot10^9 \,L_{\odot}$ \citep[e.g.][]{kade2024}, we derive a total \cii\ velocity integrated galaxy flux following \citet[]{CarilliWalter2013} \citep[see also][]{solomon1992}:
\begin{equation}
    L_{\rm [CII]} = 1.04 \cdot 10^{-3} S_{\rm [CII]} \Delta V D_{L}^2 \nu_{\rm obs} L_{\odot}, 
\end{equation}
where $D_{L}$ is the luminosity distance in Mpc, $\nu_{\rm obs}$ the observed frequency in GHz, and $S_{\rm [CII]} \Delta V$ is the total integrated \cii\ line flux in Jy~km\,s$^{-1}$, which in our case derives to $6.466$\,Jy\,km\,s$^{-1}$  ($\nu_{\rm obs} = 311.563426$\,GHz).  
In our simulations we set the integrated line flux, $S \Delta V = 6.47$\,Jy\,km\,s$^{-1}$, equal to $\Sigma_{i} S_i \Delta V$, where $\Delta V$ is the velocity step size and $S_i$ is the bin brightness. 
The total galaxy flux is conserved for all models. 
The analysis presented here can be extrapolated to any emission line observed in a high-$z$ galaxy with ALMA, {\it JWST}, or similar observatories, as long as the emission is likely to trace the galaxy disk.

\begin{table*}
\caption{Model galaxy setups. The flux, rotation velocity, and line-of-sight velocity profiles are detailed in Section \ref{ch:theoreticalgalaxyprofiles}. The first half of this table lists the parameters that are the same for all the modelled disks. The second part of the table lists each setup's abbreviated name and their specific individual settings; the 6 setups above the dotted line are set for studying the effect of the choice of moment profiles, whereas the 6 setups below the dotted line focus on the impact of rotation versus dispersion-dominated systems. }\label{table:simgals_for_simalma}
\centering
\begin{tabular}{l c c c c c c c}
\hline\hline
\noalign{\smallskip}
Fixed parameters
\\ \hline \noalign{\smallskip}
kpc/arcsec & 6.365 & & & & \\
inclination ($^\text{o}$) & 45 &  &  &  &  \\
Position angle ($^\text{o}$) & 45 & & & & \\
$r_{\rm d}$ ($''$) & 0.2 &  &  &  &  \\
Pixel size ($''$\,pix$^{-1}$)  & 0.01 & & & & \\
Spectral resolution (km/s)  & 10 & & & & \\
\hline \hline 
\noalign{\smallskip}
Abbreviation        & Flux profile & Rotation curve & Dispersion profile        & $V_{\rm max}$      & $\sigma_{\rm V}$ at $r_{\rm d}$        \\ 
        &  &  &         & [km\,s$^{-1}$]      & [km\,s$^{-1}$]        \\ 
\hline \noalign{\smallskip}
exp-risFlat-const     & exponential & rising-to-flat & constant & 250 & 50  \\ 
exp-risFlat-exp     & exponential & rising-to-flat & exponential & 250  & 50  \\ 
exp-risDec-const     & exponential & rising-to-declining & constant & 250  & 50  \\ 
exp-const-const     & exponential & constant & constant & 250  & 50  \\ 
corelog-risFlat-const     & cored-logarithmic & rising-to-flat & constant & 250  & 50  \\ 
corelog-risFlat-exp     & cored-logarithmic & rising-to-flat & exponential & 250  & 50  \\ 
\hdashline
rotDom-constDisp     & exponential & rising-to-flat & constant & 250  & 25  \\ 
rotDom-expDisp     & exponential & rising-to-flat & exponential & 250  & 25  \\ 
inBetween-constDisp     & exponential & rising-to-flat & constant & 150  & 150  \\ 
inBetween-expDisp     & exponential & rising-to-flat & exponential & 150  & 150  \\ 
dispDom-constDisp     & exponential & rising-to-flat & constant & 50  & 250  \\ 
dispDom-expDisp     & exponential & rising-to-flat & exponential & 50  & 250  \\ 
\hline 
\hline  
\end{tabular}
\end{table*}

\begin{table*}
\centering
\caption{The settings for the ALMA synthetic observations created using \texttt{simalma}. All ALMA synthetic observations are carried out on the sample model galaxy setups listed in Table \ref{table:simgals_for_simalma}. Integration time, precipitable water vapour, ALMA configuration, and pixel size are set in the synthetic observation process. The SNR and beam size are derived and obtained from the resultant synthetically observed cubes.}\label{table:intrinsicNobsALMA}
\begin{threeparttable}
\begin{tabular}{l c c c c c c c c}
\hline\hline
\noalign{\smallskip}
\multicolumn{3}{l}{Synthetic ALMA observations}&  &  & & \\ \hline \hline 
                            & SNR\tnote{a} & Integration time & pwv\tnote{b}        & ALMA config      & pixel size        & Resultant beam size & Beam PA\\ 
           Data quality     &                       & (s)               & (mm)            &                   & ($''\times''$)     &   ($''\times''$) [kpc$\times$kpc] & (deg)\\ \hline 
              High          &  20                   & 21600             &  0.472        &    8.6            & 0.02$\times$0.02     & 0.16$\times$0.12 [1.02$\times$0.76]  & -84                 \\
          Medium            &  15                   & 7200              &  0.7855       &    8.4            & 0.04$\times$0.04   &   0.40$\times$0.36 [2.55$\times$2.29]  & -79            \\
          Low               &  9                    & 3400              &  1.262        &    8.2            & 0.1$\times$0.1   &   1.03$\times$0.87 [6.56$\times$5.54]    & -88                  \\ \hline 
\hline  
\end{tabular}
\begin{tablenotes}
\small
\item[a] Signal-to-noise ratio is derived as the peak value in the central pixel of the galaxy divided by the standard deviation of the noise measured in an emission-free area of the cube.
\item[b] Best weather conditions at 0.472, first octile, 0.7855 in between second and third octile, and fourth octile 1.262.
\end{tablenotes}
\end{threeparttable}
\end{table*}


\section{Synthetic ALMA observations}\label{ch:simalma}
The simulated synthetic ALMA observations were carried out using CASA, \texttt{simalma} and \texttt{tclean} \citep{casa}. 
For the \texttt{simalma} observations, we set the integration time, the precipitable water vapour (pwv) level, and the ALMA configuration according to Table \ref{table:intrinsicNobsALMA}. 
The \texttt{simalma} synthetically observed cubes were cleaned using tclean and natural weighting. 
The data cubes were spatially and spectrally binned in the cleaning process, spatially going from a pixel size of $0.01''\times0.01''$ to the respective values listed in Table \ref{table:intrinsicNobsALMA} and spectrally binned by a factor of 2 to a spectral channel width of 20 km\,s$^{-1}$. 
The pixel size for the synthetic ALMA observations are set to a minimum of 5 pixels per beam minor axis, to ensure accurate reproduction of the signal in the resultant cube. 
Each model disk was synthetically observed at three different combinations of integration time, pwv, ALMA configuration, and pixel size to achieve three different levels of resultant data quality: high, medium, and low, as listed in the Table \ref{table:intrinsicNobsALMA}. 
The number of independent resolution elements across the synthetically observed disks are derived using the beam sizes presented in Table \ref{table:intrinsicNobsALMA} at a SNR cutoff of 3, and depend on the chosen intrinsic brightness profile and the dispersion, due to the conserved total flux independent of intrinsic model.

It is worth noting that while we do not recommend the low data quality settings to be used for resolved galaxy kinematics studies, this kind of setup is sometimes used and it is therefore important to include in the analysis.

\section{3D-kinematical fitting tools and setup}
\label{ch:3Dfitting}  
Our disk models are fitted with three different 3D-kinematical tools: \barolo\, \citep{diteodoro2015}, \galpak\, \citep{galpakcode2015,bouche2015}, and \qubefit\, \citep{neeleman2020thecode,neeleman2021}, which were developed to handle high-redshift/low-resolution galaxies, and are often used and publicly available. 
All three tools assume axisymmetric thin rotating disks. 
\barolo\, relies on tilted rings to optimise a model galaxy disk \citep{rogstad1974,bouche2015}, each ring is fitted individually. 
\galpak\, and \qubefit\, adopt parametric models, i.e., pre-fit chosen equations describing the brightness, rotation velocity, and velocity dispersion as a function of radius. These tools optimise the kinematical parameters of the equations describing the model disk. 

The degree of freedom, i.e., the number of free parameters for each tool, includes: \\
for \barolo, for each ring: rotation velocity, velocity dispersion, position angle and or inclination depending on the specific run; \\
for \galpak: maximum rotation velocity, velocity dispersion, inclination and or position angle depending on the specific run, turnover radius, flux, maximum radius; \\
for \qubefit\, : maximum rotation velocity, velocity dispersion, inclination and or position angle depending on the specific run, turnover radius, flux.

For the \barolo\, fits, we set the disk thickness to less than the pixel size, setting the normalisation to azimuthal and the masking algorithm to \texttt{smooth\&search}, with the exception of the low-quality data where we use the \texttt{search} masking method due to the poorly resolved data. 
We note that, \barolo modelling performance does not qualitatively change if we consider different choices for the SNR cut-off limit. In Appendix~\ref{app:barolo_SNR_section} we show that this holds for both the recovery of the rotation velocity and velocity dispersion (but see \citealt{lee2024}). 
Furthermore, the number of rings remain consistent independent of the SNR cut-off value and the \barolo\, default of SNR cut-off of 5 and smoothing factor 2 provides the best masks with the least amount of excess noisy regions and most accurate extent relative to the number of rings. 
The \barolo\, ring width is set to 1/2.5 of the beam major axis FWHM and the number of rings is set to the highest number possible while still ensuring that even the outermost ring is completely filled. 

For the parametric models, \galpak\, and \qubefit, we choose brightness and rotation velocity profiles that are similar and comparable between the two tools, resulting in a standard setting of an exponential flux profile and an arctan rotation curve. 
Regarding the velocity dispersion, for \galpak\, our standard dispersion setup is a thin disk, as our models are infinitesimally thin and it enables a more accurate comparison to \qubefit, where we use the thin disk model. 
Nevertheless, we also study the effect of assuming thick-disk geometry in the modelling performance. 
For \qubefit\,  the standard velocity dispersion profile is a constant profile, but we also investigate the effect of an exponential dispersion profile, as well as a constant rotation curve. 
For \galpak\, and \qubefit\, we adopt the defaults of no masking and a SNR cut-off of 2 respectively. 
As far as possible, we adopt the same start values for each fit independent of tool. 

To better emulate the realistic process and results obtained from real observational data, we make sure to vary the free parameters, while the flux, rotation velocity, and velocity dispersion are always free (and the scale radius when applicable), we vary combinations of free inclination and/or free position angle, as well as investigate the effect of an incorrect assumed fixed inclination (this option is separated in the result section from the others). 
For systems with expected non-axisymmetric structures \barolo\, fits are in some cases run twice, once with free position angle and inclination and a second time with the averages of the position angle and inclinations obtained in the first run. Our disks do not include non-axisymmetric structures and the run with the fixed position angle and fixed inclination to their respective correct values can be taken as a substitute for this fitting method. 
The full list of the profile settings and free parameters used for the 3D fitting is listed in Table \ref{table:3Dconvergence_barolo}-\ref{table:3Dconvergence_qubefit} together with a note on whether the fit has converged or not. 

Furthermore, note that while \barolo, \galpak, and \qubefit\, all present modelled maximum rotation and velocity dispersion values, these values are not derived or presented in the same manner. 
In general, when comparing kinematical values of galaxies caution is advised as the maximum rotation velocity can be the velocity at the maximum radius of the field-of-view of the cube, at $\sim2.2 r_{\rm d}$, at the optical radius, at the effective radius, or as an average where the rotation curve is flat (if it becomes flat), whereas the velocity dispersion can be the turbulent non-disk term, the combined disk and turbulent dispersion, or the combined disk, turbulent, and instrumentally broadened dispersion. 
In this work, to ensure a correct comparison between three tools and to avoid issues relating to how each of them define the parameters, we thereby work only with the 3D-kinematical tools' output model cubes, and any parameter values we present in this paper have been extracted directly from the cubes themselves at radii or manner stated.

\section{Results}\label{ch:results}
We extract the rotation velocity and velocity dispersion values from the intrinsic, synthetically observed, and model cube in three different ways for each: 
\begin{itemize}
    \item The de-projected rotation velocity value extracted 
    \begin{itemize}
        \item at $1r_{\rm d}$, and at $2.2r_{\rm d}$,
        \item and as the maximum value obtained within a radius large enough to include the full synthetically observed signal and thereby the full extent of the region modelled (radius of $1''$).
    \end{itemize}
    \item The velocity dispersion value, 
    \begin{itemize}
        \item at $1r_{\rm d}$, and at $2.2r_{\rm d}$, 
        \item and as the mean value within the $1''$ radius ($\bar{\sigma_{\rm V}}$). 
    \end{itemize}
\end{itemize}
These values are derived from the extraction of a one-beam wide slit along the major axis of the galaxy, and the $1r_{\rm d}$, and at $2.2r_{\rm d}$ values are calculated as an average in a 10\% range around the 1 and 2.2 scale radius, respectively. 
As the line emission seen in the 3D-fitted model cubes and the intrinsic cubes cover the full field-of-view, exceptions for missing values or large value changes are not necessary. 
But for the synthetically observed cube, the cube is masked at a SNR of 3 and suffers from noise, so for the $2.2r_{\rm d}$ radius, we ensure that the highest radial values are not masked out and that they do not suffer from edge effects by checking that the change between the second to last data point and the last data point is smaller than 10\%, which is a suitable assumption for both rotation velocities and velocity dispersions in the outskirts of galaxies. 
The de-projected rotation velocity value is calculated using the 3D fitted model inclination for the model cube and the intrinsic value for the intrinsic ideal cube and the synthetically observed cube. 
We avoid comparisons using the number of standard deviations offset between the 3D fit model and the intrinsic, as that would unfairly bias towards fits with large error bars; we instead compare the model best fit value against the intrinsic corresponding value. 

\begin{figure*}
    \centering
    \includegraphics[width=\textwidth]{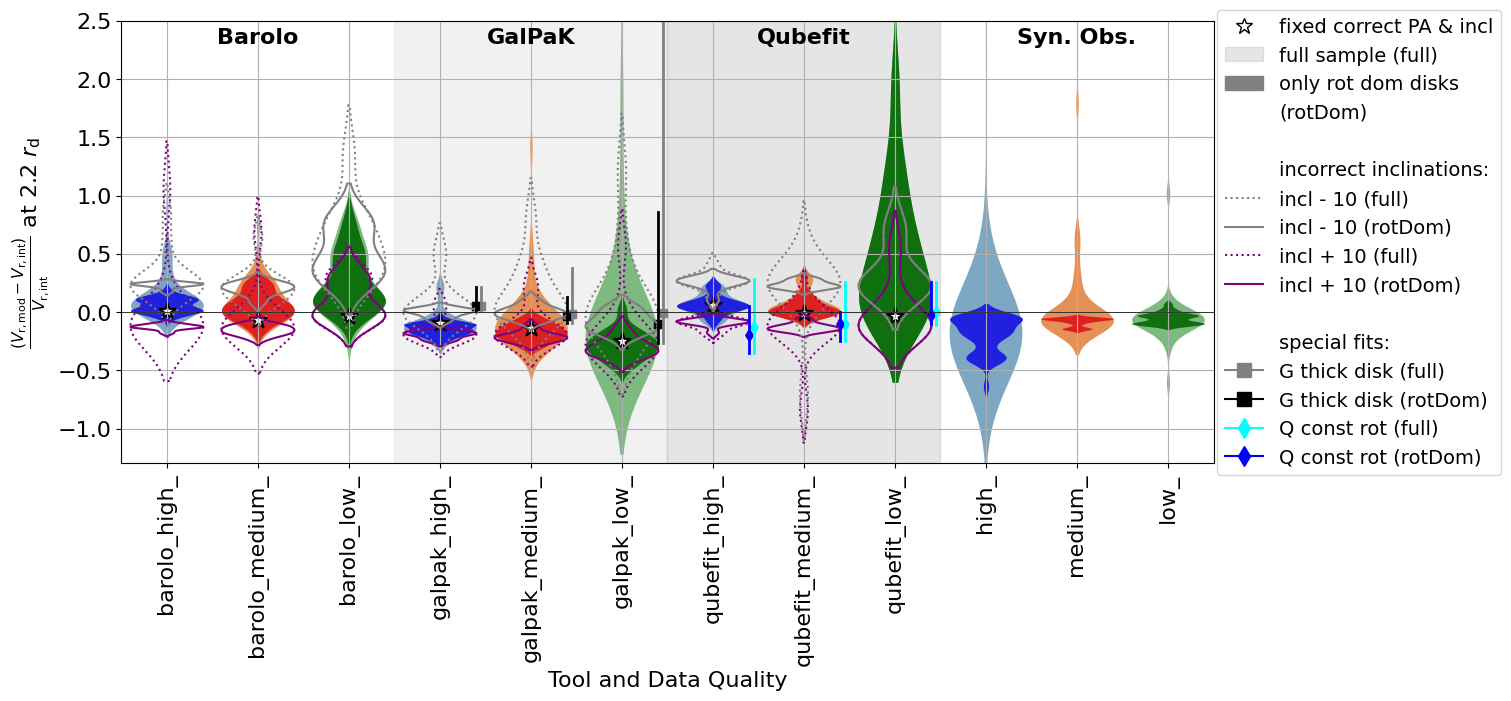}
    \caption{The offset of the de-projected rotation velocity from the intrinsic value for each data quality and 3D-kinematical tool's converged models in the first three parts of the plot, while the fourth and right-most part displays the offset between the synthetically observed cube and the intrinsic. The transparent colouring indicates the extent of the full dataset, including rotation-dominated, liminal, and dispersion-dominated systems, whereas the opaque colours show only the rotation-dominated disks fitted with a range of combinations of correctly fixed and/or free position angle and/or inclination. 
    The white stars pinpoint the values of the fits with intrinsically correct position angle and inclination (at an exponential brightness profile, an arctan rotation curve and a constant dispersion profile for the parametric tools). 
    The light (grey) outline (and dotted line) indicates the extent of the position angle offset of the rotation-dominated systems (full disk sample) fitted with a fixed inclination set to 10 degrees below the intrinsic value, whereas the dark (purple) outline (and dotted line) shows the offset for the rotation-dominated systems (full disk sample) fitted with a fixed inclination at 10 degrees above the intrinsic value. 
    The black (and grey) lines next to the \galpak\, violins illustrate the results for the rotation-dominated systems (full disk sample) fitted with a thick disk instead of a thin one. The dark (blue) and light (cyan) lines next to the \qubefit\, violins denote the fit parameter results of the rotation-dominated systems (full disk sample) when fitted with a constant rotation curve instead of the default arctan profile. 
    The x-axis shows the combination of tool --\barolo, \galpak, or \qubefit-- together with the data quality: high, medium, or low. 
    Offset and scatter values are presented in Table \ref{tab:fit_eval_params}.}
    \label{fig:violin_Vrot_22rturn}
\end{figure*}

\begin{figure*}
    \centering
    \includegraphics[width=\textwidth]{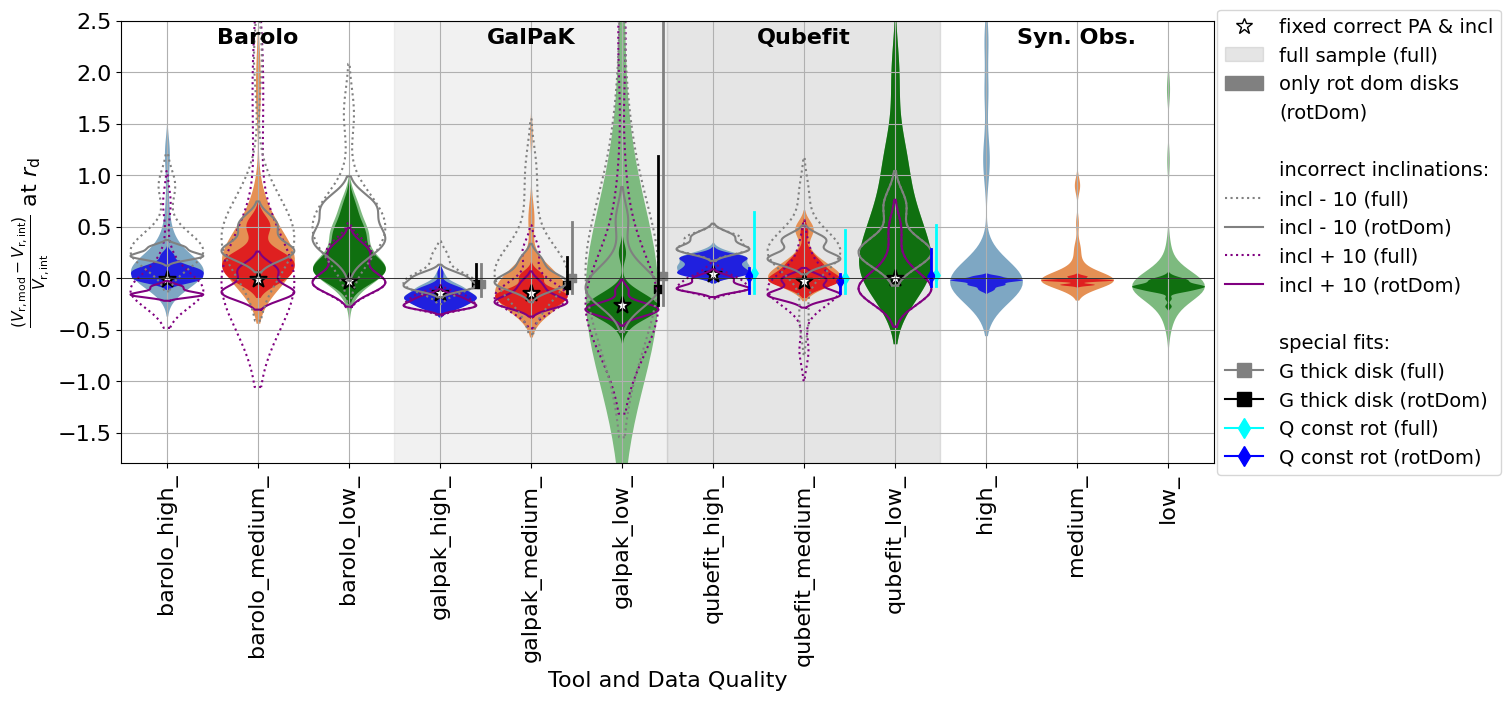}
    \caption{The offset of the de-projected rotation velocity from the intrinsic value at the scale radius. The figure is colour coded as Fig.~\ref{fig:violin_Vrot_22rturn}.
    }
    \label{fig:violin_Vrot_rturn}
\end{figure*}

\begin{figure*}[h]
    \centering
    \includegraphics[width=\textwidth]{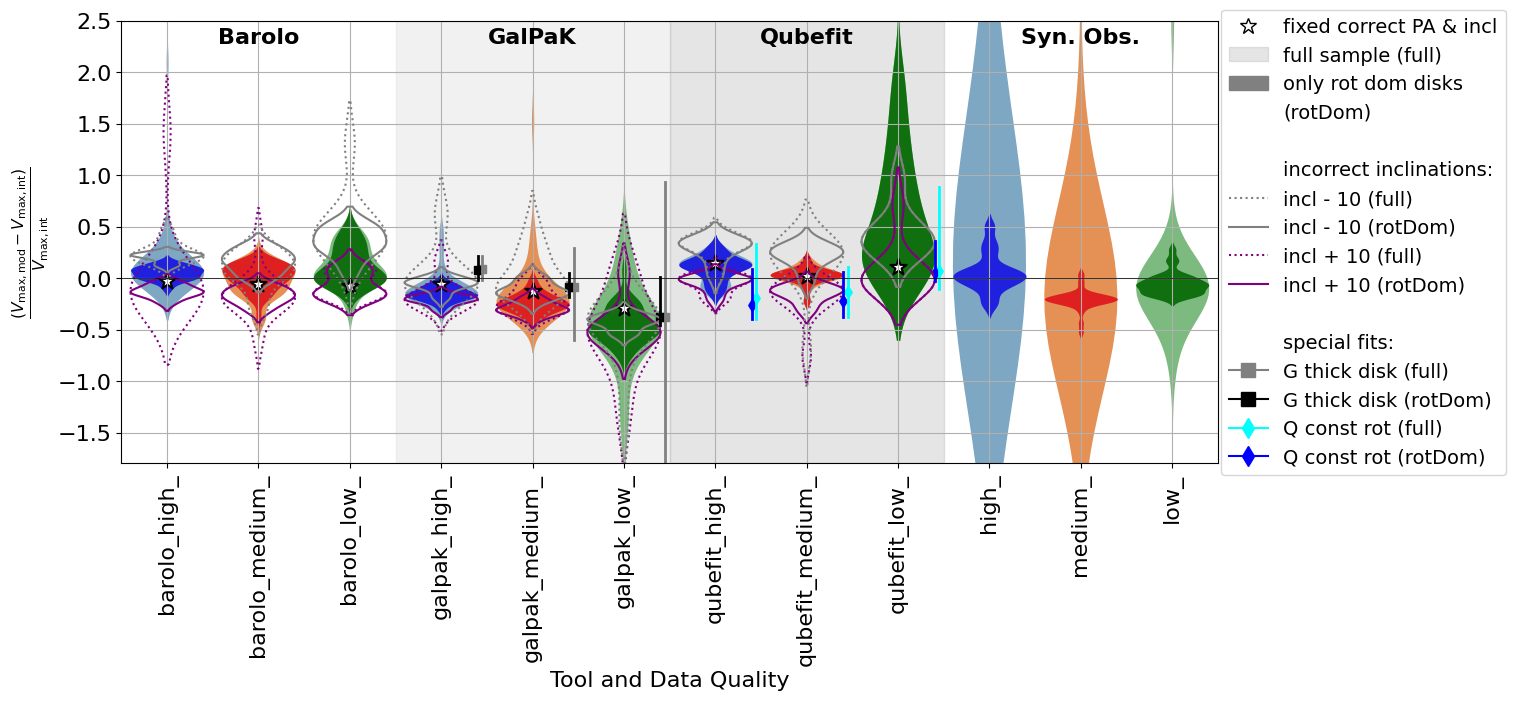}
    \caption{The offset of the maximum de-projected rotation velocity from the maximum intrinsic value along the major. The figure is colour coded as Fig.~\ref{fig:violin_Vrot_22rturn}. 
    }
    \label{fig:violin_Vrot_max}
\end{figure*}

\begin{figure*}[h]
    \centering
    \includegraphics[width=\textwidth]{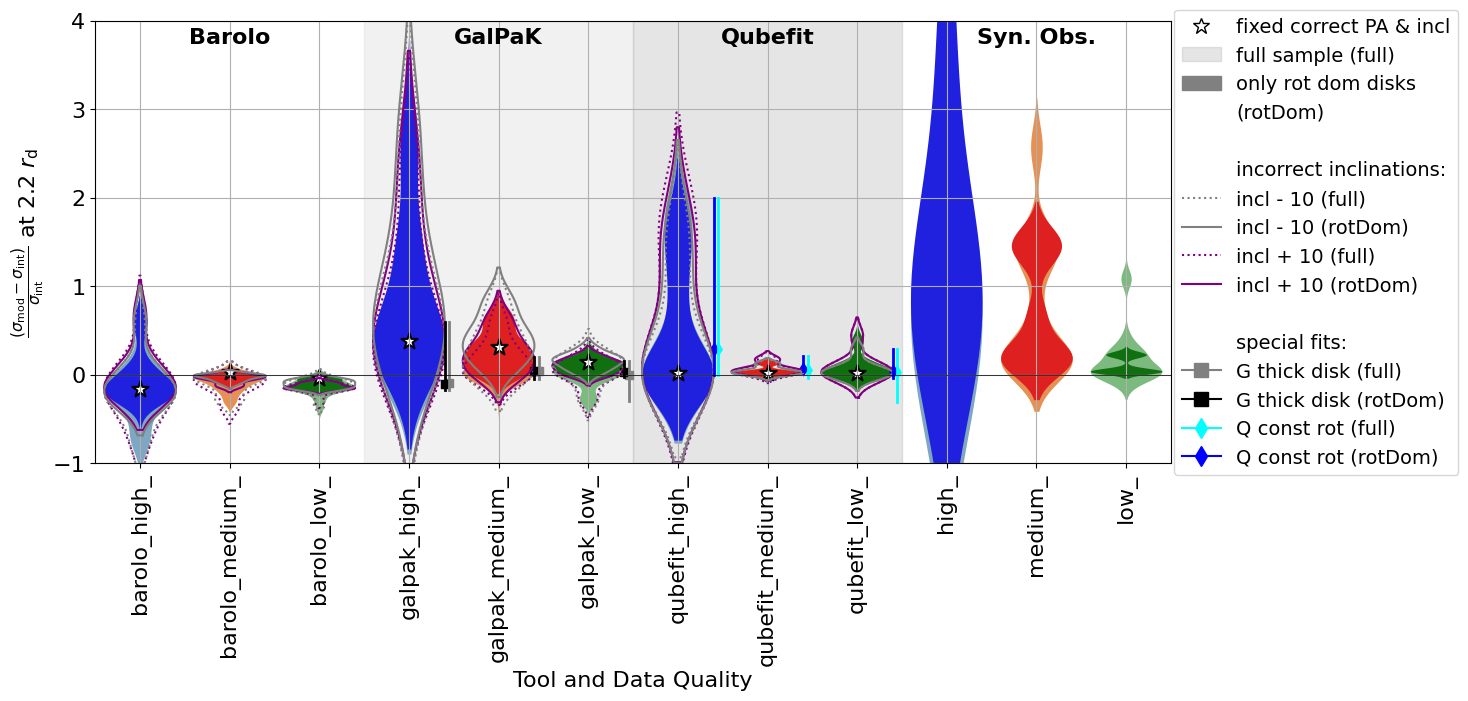}
    \caption{The offset of the velocity dispersion from the intrinsic value at 2.2$\times$ the scale radius. The figure is colour coded as Fig.~\ref{fig:violin_Vrot_22rturn}.
    .}
    \label{fig:violin_Vdisp_22rturn}
\end{figure*}

\begin{figure*}[h]
    \centering
    \includegraphics[width=\textwidth]{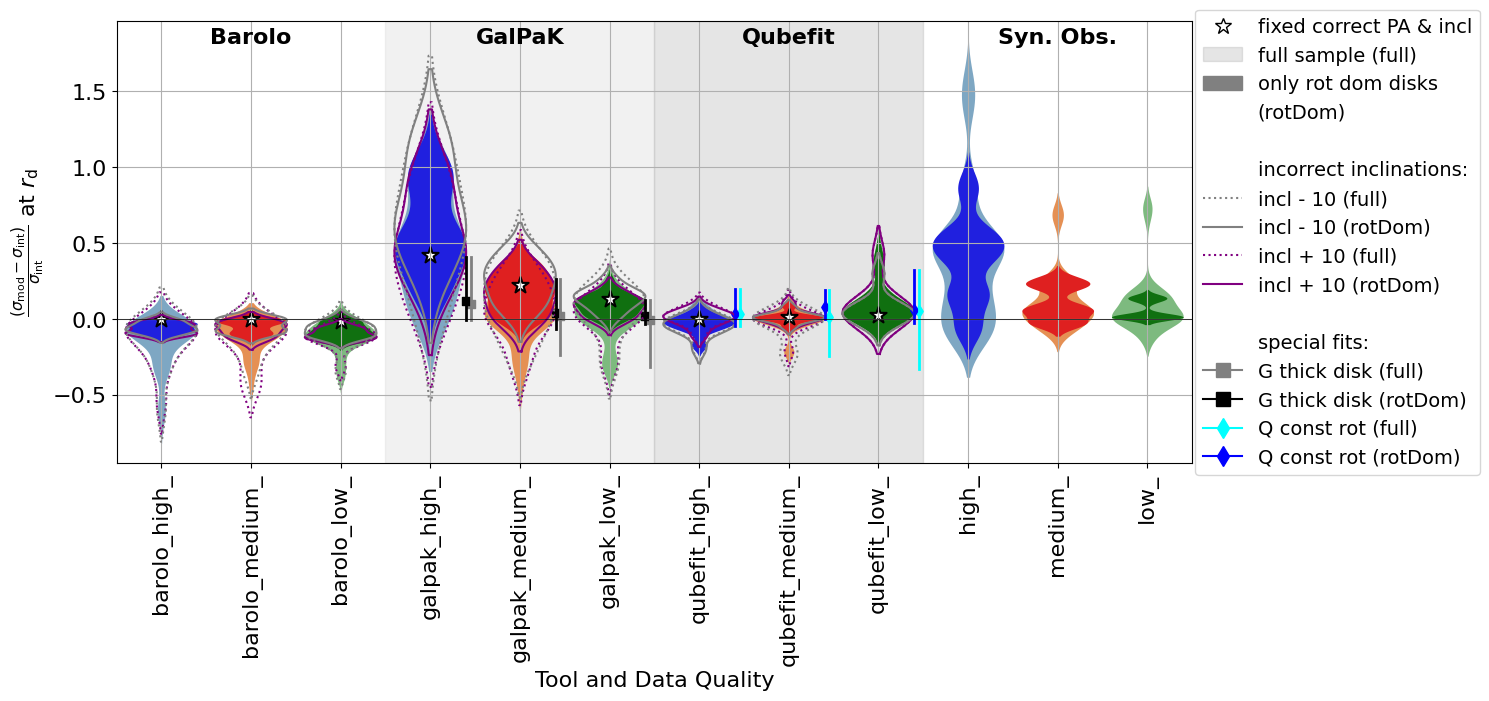}
    \caption{The offset of the velocity dispersion from the intrinsic value at the scale radius.The figure is colour coded as Fig.~\ref{fig:violin_Vrot_22rturn}. 
    }
    \label{fig:violin_Vdisp_rturn}
\end{figure*}

\begin{figure*}[h]
    \centering
    \includegraphics[width=\textwidth]{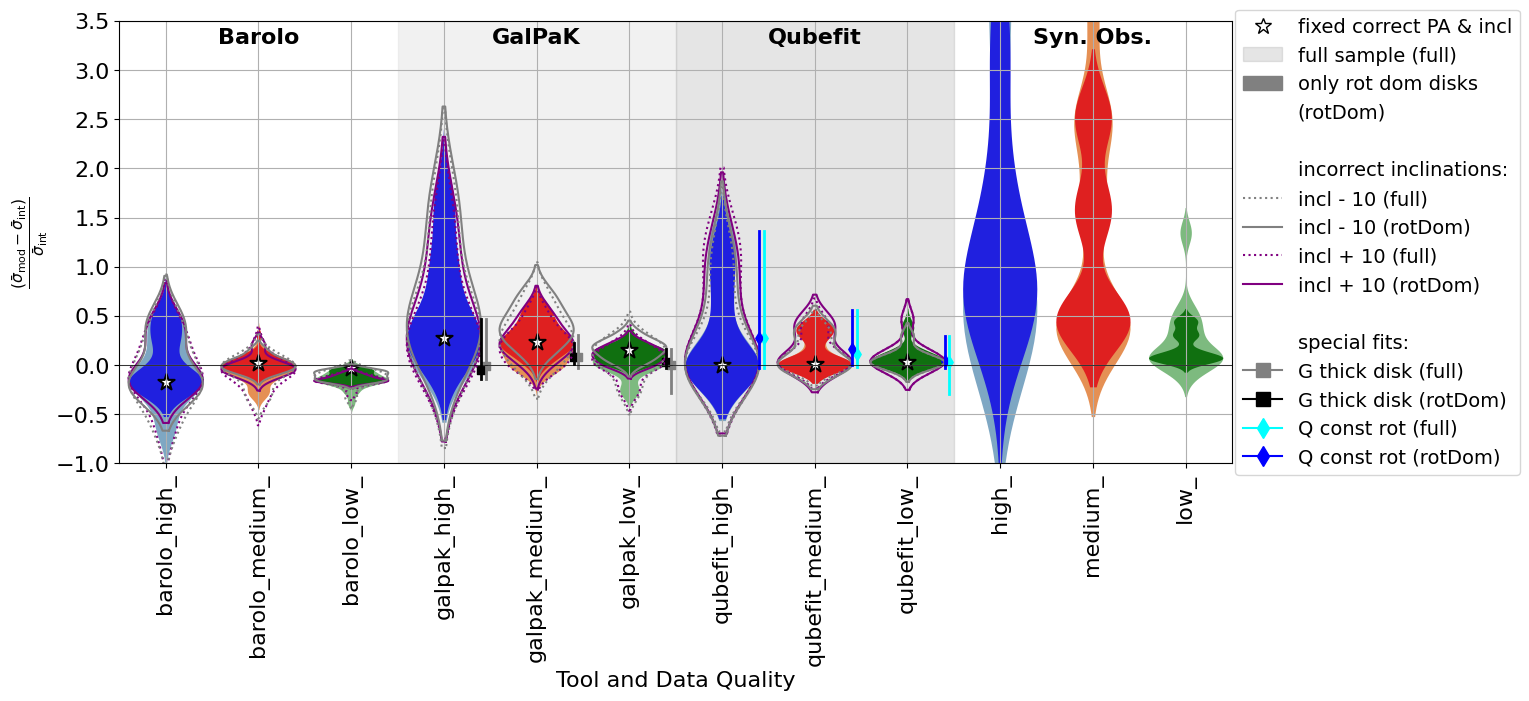}
    \caption{The offset of the velocity dispersion from the intrinsic value averaged across the major axis. The figure is colour coded as Fig.~\ref{fig:violin_Vrot_22rturn}. 
    }
    \label{fig:violin_Vdisp_mean}
\end{figure*}

The rotation velocity and velocity dispersion values of the converged systems (the list of which systems have converged is found in Appendix \ref{app:tables:3Dconvergence}) are presented on the form: 
$(model\_value-intrinsic\_value)/intrinsic\_value$ in Fig.~\ref{fig:violin_Vrot_22rturn}-\ref{fig:violin_Vdisp_mean}, and as the percentage in Table \ref{tab:fit_eval_params}. 
The fourth section of the figures shows the parameters offsets and scatters as extracted directly from the synthetically observed cubes masked at a SNR of 3. 
The plots in the figures are zoomed in to facilitate visual interpretation, the minimum and maximum values for each parameter range are listed in Table \ref{tab:fit_eval_params} for the three 3D-kinematical tools. 

In general, the three tools show difficulties with converging for dispersion-dominated systems($V_{\rm max}/\sigma_{\rm V} < 1$), \barolo\, for low data quality dispersion-dominated systems, \galpak\, for high data quality dispersion-dominated systems, and \qubefit\, for all data quality dispersion-dominated systems as well as in-between systems ($V_{\rm max}/\sigma_{\rm V} \approx 1$) at low data quality. 
On average, the convergence rate for the three tools for rotation dominated systems is 100\%, 92\%, and 99\% for \barolo, \galpak, and \qubefit\, respectively. 
For non-ideal disks ($V_{\rm rot}/\sigma_{\rm V}\leq 1$ the convergence rate is 81\%, 73\%, and 46\% respectively.

In Fig.~\ref{fig:violin_Vrot_22rturn}-\ref{fig:violin_Vdisp_mean}, the darkest violins are the kinematical parameter accuracy of the optimal cases where the intrinsic system is rotation dominated ($V_{\rm max}/\sigma_{\rm V} = 5, 10$) and fitted with a range of combinations of correctly fixed or free position angles and/or inclination, and, when applicable, fitted with an arctan rotation curve and a constant dispersion profile (section \ref{ch:3Dfit_results_rot_dom}). 
The white star highlights the fit where the position angle and inclination are fixed to the intrinsically correct values, and, in the case of the parametric tools, the profile settings are exponential brightness profile, arctan rotation curve, and constant dispersion profile. 
The transparent violins include both rotation dominated ($V_{\rm max}/\sigma_{\rm V} = 5, 10$), dispersion dominated ($V_{\rm max}/\sigma_{\rm V} = 0.2$), and in-between systems ($V_{\rm max}/\sigma_{\rm V} = 1$), illustrating the range you can expect when you do not know whether your galaxy is rotation dominated or not (section \ref{ch:3D_fit_param_disp_dom}). 
The light (grey) and dark (purple) outlines show the effect of fitting with an assumed inclination that is 10 degrees too large or 10 degrees too small, this is detailed further in section \ref{ch:3D_fit_section_incl10}. 
The dark (black) and light (grey) lines next to the \galpak\, violins denote the fits with a thick disk instead of the default thin disk, detailed in section \ref{ch:galpak_thick}. 
The dark (blue) and light (cyan) lines next to the \qubefit\, violins illustrate the results of the fits with a constant rotation curve, presented in section \ref{ch:qubefit_const_rot}. 
And in section \ref{ch:qubefit_expdisp_or_const} we explore the effect of fitting galaxy disks with intrinsic constant or exponential dispersion profiles using \qubefit\, modelled constant or exponential dispersion profiles.

\setlength{\tabcolsep}{1pt}
\begin{table*}
\scriptsize
    \caption{Kinematic parameter deviation from the intrinsic values as recovered by \barolo, \galpak, and \qubefit\, converged models. For each parameter deviation, we present the median, $16^{\rm th}$, and $84^{\rm th}$ percentile values of the distribution in percentages, plus the minimum and maximum percentage values inside brackets (Median$^{rel.84\%}_{rel.16\%}[min;max]$). 
    ``(All)'' denotes the sample containing both dispersion-dominated, liminal, and rotation-dominated disks, whereas the right-hand values, ``(RotDom)'', in blue are for the rotation-dominated disks ($V/\sigma_{\rm V} > 1$) only.
    }
    \label{tab:fit_eval_params}
    \centering
\begin{tabular}{lcccccccc} \hline  \hline 
                    &              &             &            &  \\
              \textbf{\barolo} & \multicolumn{2}{c}{High data quality} & \multicolumn{2}{c}{Medium data quality} & \multicolumn{2}{c}{Low data quality} & \multicolumn{2}{c}{All data qualities} \\
          Parameter & (All) & {\color{blue}(Rot Dom)} & (All) & {\color{blue}(Rot Dom)} & (All) & {\color{blue}(Rot Dom)} & (All) & {\color{blue}(Rot Dom)} \\ \hline  \smallskip 
$V$ at $2.2r_{\rm d}$ (\%) & $9^{+19}_{-9}[-5;88]$ &\textcolor{blue}{4$^{+8}_{-3}[-5;19]$} & $5^{+20}_{-10}[-15;75]$ &\textcolor{blue}{5$^{+16}_{-10}[-15;26]$} & $15^{+36}_{-10}[-20;82]$ &\textcolor{blue}{15$^{+35}_{-10}[-3;82]$} & $10^{+29}_{-11}[-20;88]$ &\textcolor{blue}{8$^{+15}_{-9}[-15;82]$} \\ \smallskip 
$V$ at r$_{\rm d}$ (\%) & $9^{+22}_{-10}[-9;126]$ &\textcolor{blue}{6$^{+6}_{-8}[-3;24]$} & $20^{+34}_{-18}[-1;213]$ &\textcolor{blue}{16$^{+29}_{-15}[-1;56]$} & $15^{+30}_{-13}[-20;79]$ &\textcolor{blue}{15$^{+30}_{-12}[-2;76]$} & $15^{+33}_{-14}[-20;213]$ &\textcolor{blue}{12$^{+23}_{-12}[-3;76]$} \\ \bigskip 
$V_\text{rot,max}$ (\%) & $7^{+10}_{-10}[-11;209]$ &\textcolor{blue}{6$^{+6}_{-7}[-11;16]$} & $-1^{+14}_{-15}[-44;25]$ &\textcolor{blue}{5$^{+8}_{-18}[-37;18]$} & $7^{+28}_{-12}[-33;52]$ &\textcolor{blue}{5$^{+29}_{-9}[-17;52]$} & $5^{+15}_{-14}[-44;209]$ &\textcolor{blue}{5$^{+10}_{-10}[-37;52]$} \\ \smallskip 
$\sigma_{\rm V}$ at $2.2r_{\rm d}$ (\%) & $-18^{+23}_{-32}[-82;62]$ &\textcolor{blue}{-15$^{+23}_{-7}[-34;62]$} & $-5^{+5}_{-14}[-33;7]$ &\textcolor{blue}{-1$^{+2}_{-5}[-11;7]$} & $-14^{+10}_{-4}[-39;-1]$ &\textcolor{blue}{-11$^{+7}_{-5}[-18;-1]$} & $-12^{+12}_{-14}[-82;62]$ &\textcolor{blue}{-7$^{+8}_{-10}[-34;62]$} \\ \smallskip 
$\sigma_{\rm V}$ at r$_{\rm d}$ (\%) & $-8^{+6}_{-22}[-66;0]$ &\textcolor{blue}{-5$^{+3}_{-3}[-12;0]$} & $-9^{+8}_{-13}[-46;0]$ &\textcolor{blue}{-3$^{+3}_{-7}[-13;0]$} & $-11^{+9}_{-4}[-39;3]$ &\textcolor{blue}{-8$^{+7}_{-5}[-15;3]$} & $-9^{+7}_{-17}[-66;3]$ &\textcolor{blue}{-6$^{+5}_{-6}[-15;3]$} \\ \smallskip
$\bar{\sigma_{\rm V}}$ (\%) & $-13^{+49}_{-18}[-71;58]$ &\textcolor{blue}{-12$^{+52}_{-8}[-24;58]$} & $-4^{+10}_{-13}[-33;31]$ &\textcolor{blue}{1$^{+7}_{-7}[-11;31]$} & $-15^{+9}_{-4}[-40;-2]$ &\textcolor{blue}{-12$^{+7}_{-4}[-19;-2]$} & $-11^{+18}_{-13}[-71;58]$ &\textcolor{blue}{-7$^{+17}_{-10}[-24;58]$} \\
\hline\hline

        &              &             &            &  \\
  \textbf{\galpak} & \multicolumn{2}{c}{High data quality} & \multicolumn{2}{c}{Medium data quality} & \multicolumn{2}{c}{Low data quality} & \multicolumn{2}{c}{All data qualities} \\
Parameter & (All) & {\color{blue}(Rot Dom)} & (All) &  {\color{blue}(Rot Dom)} & (All) & {\color{blue}(Rot Dom)} & (All) &  {\color{blue}(Rot Dom)} \\ \hline  \smallskip 
$V$ at $2.2r_{\rm d}$ (\%) & $-11^{+4}_{-7}[-24;23]$ &\textcolor{blue}{-15$^{+5}_{-4}[-24;-4]$} & $-13^{+18}_{-10}[-30;140]$ &\textcolor{blue}{-19$^{+8}_{-5}[-30;-3]$} & $-21^{+25}_{-14}[-48;446]$ &\textcolor{blue}{-28$^{+11}_{-12}[-48;3]$} & $-15^{+18}_{-13}[-48;446]$ &\textcolor{blue}{-19$^{+8}_{-12}[-48;3]$} \\ \smallskip 
$V$ at r$_{\rm d}$ (\%) & $-18^{+10}_{-7}[-29;5]$ &\textcolor{blue}{-18$^{+8}_{-8}[-29;5]$} & $-13^{+19}_{-11}[-29;141]$ &\textcolor{blue}{-17$^{+14}_{-9}[-29;6]$} & $-19^{+83}_{-14}[-47;1032]$ &\textcolor{blue}{-25$^{+12}_{-10}[-47;28]$} & $-17^{+24}_{-11}[-47;1032]$ &\textcolor{blue}{-20$^{+9}_{-9}[-47;28]$} \\ \bigskip
$V_\text{rot,max}$ (\%) & $-15^{+12}_{-5}[-30;46]$ &\textcolor{blue}{-15$^{+7}_{-6}[-30;-1]$} & $-21^{+19}_{-12}[-43;153]$ &\textcolor{blue}{-25$^{+9}_{-8}[-43;0]$} & $-47^{+37}_{-22}[-209;50]$ &\textcolor{blue}{-48$^{+23}_{-12}[-77;8]$} & $-23^{+20}_{-26}[-209;153]$ &\textcolor{blue}{-25$^{+11}_{-24}[-77;8]$} \\ \smallskip 
$\sigma_{\rm V}$ at $2.2r_{\rm d}$ (\%) & $40^{+105}_{-31}[0;292]$ &\textcolor{blue}{48$^{+114}_{-16}[9;292]$} & $11^{+35}_{-13}[-25;73]$ &\textcolor{blue}{27$^{+33}_{-22}[1;73]$} & $4^{+13}_{-9}[-30;30]$ &\textcolor{blue}{11$^{+8}_{-8}[-2;30]$} & $11^{+37}_{-13}[-30;292]$ &\textcolor{blue}{20$^{+39}_{-15}[-2;292]$} \\ \smallskip 
$\sigma_{\rm V}$ at r$_{\rm d}$ (\%) & $48^{+44}_{-31}[0;105]$ &\textcolor{blue}{52$^{+42}_{-30}[15;105]$} & $7^{+22}_{-9}[-46;40]$ &\textcolor{blue}{19$^{+14}_{-17}[-2;40]$} & $4^{+10}_{-9}[-32;26]$ &\textcolor{blue}{10$^{+6}_{-7}[-3;26]$} & $9^{+30}_{-11}[-46;105]$ &\textcolor{blue}{18$^{+31}_{-14}[-3;105]$} \\ \smallskip 
$\bar{\sigma_{\rm V}}$ (\%) & $29^{+75}_{-26}[-1;182]$ &\textcolor{blue}{38$^{+75}_{-16}[3;182]$} & $19^{+22}_{-21}[-12;64]$ &\textcolor{blue}{25$^{+26}_{-11}[1;64]$} & $3^{+14}_{-8}[-29;32]$ &\textcolor{blue}{11$^{+9}_{-8}[-3;32]$} & $14^{+26}_{-16}[-29;182]$ &\textcolor{blue}{22$^{+28}_{-17}[-3;182]$} \\
\hline\hline

        &              &             &            &  \\
  \textbf{\qubefit} & \multicolumn{2}{c}{High data quality} & \multicolumn{2}{c}{Medium data quality} & \multicolumn{2}{c}{Low data quality} & \multicolumn{2}{c}{All data qualities} \\
Parameter & (All) &  {\color{blue}(Rot Dom)} & (All) &  {\color{blue}(Rot Dom)} & (All) &  {\color{blue}(Rot Dom)} & (All) &  {\color{blue}(Rot Dom)} \\ \hline         \smallskip 
$V$ at $2.2r_{\rm d}$ (\%) & $5^{+11}_{-9}[-8;23]$ &\textcolor{blue}{5$^{+7}_{-4}[-8;23]$} & $4^{+14}_{-5}[-3;31]$ &\textcolor{blue}{1$^{+11}_{-3}[-3;31]$} & $34^{+68}_{-33}[-3;206]$ &\textcolor{blue}{34$^{+68}_{-33}[-3;206]$} & $5^{+26}_{-7}[-8;206]$ &\textcolor{blue}{5$^{+29}_{-6}[-8;206]$} \\ \smallskip 
$V$ at r$_{\rm d}$ (\%) & $13^{+8}_{-8}[3;28]$ &\textcolor{blue}{12$^{+9}_{-8}[3;24]$} & $6^{+21}_{-8}[-5;50]$ &\textcolor{blue}{2$^{+16}_{-6}[-5;46]$} & $38^{+68}_{-35}[-7;210]$ &\textcolor{blue}{38$^{+68}_{-35}[-7;210]$} & $12^{+34}_{-12}[-7;210]$ &\textcolor{blue}{12$^{+34}_{-12}[-7;210]$} \\ \bigskip
$V_\text{rot,max}$ (\%) & $12^{+9}_{-20}[-14;30]$ &\textcolor{blue}{12$^{+9}_{-15}[-14;30]$} & $2^{+7}_{-6}[-22;19]$ &\textcolor{blue}{2$^{+5}_{-4}[-22;19]$} & $36^{+89}_{-25}[-2;203]$ &\textcolor{blue}{36$^{+89}_{-25}[-2;203]$} & $11^{+19}_{-13}[-22;203]$ &\textcolor{blue}{11$^{+28}_{-10}[-22;203]$} \\ \smallskip 
$\sigma_{\rm V}$ at $2.2r_{\rm d}$ (\%) & $2^{+169}_{-3}[-4;192]$ &\textcolor{blue}{2$^{+139}_{-3}[-4;176]$} & $4^{+9}_{-2}[-2;14]$ &\textcolor{blue}{4$^{+10}_{-1}[1;14]$} & $4^{+10}_{-5}[-5;45]$ &\textcolor{blue}{4$^{+10}_{-5}[-5;45]$} & $3^{+33}_{-4}[-5;192]$ &\textcolor{blue}{3$^{+29}_{-4}[-5;176]$} \\ \smallskip 
$\sigma_{\rm V}$ at r$_{\rm d}$ (\%) & $-1^{+5}_{-6}[-18;10]$ &\textcolor{blue}{$-3^{+3}_{-5}[-18;5]$} & $0^{+3}_{-5}[-22;9]$ &\textcolor{blue}{1$^{+2}_{-1}[-5;9]$} & $6^{+11}_{-4}[-4;47]$ &\textcolor{blue}{6$^{+11}_{-4}[-4;47]$} & $0^{+8}_{-5}[-22;47]$ &\textcolor{blue}{1$^{+7}_{-5}[-18;47]$} \\ \smallskip 
$\bar{\sigma_{\rm V}}$ (\%) & $1^{+120}_{-3}[-6;129]$ &\textcolor{blue}{0$^{+92}_{-3}[-6;124]$} & $2^{+34}_{-4}[-2;40]$ &\textcolor{blue}{2$^{+36}_{-1}[-2;40]$} & $5^{+9}_{-5}[-4;45]$ &\textcolor{blue}{5$^{+9}_{-5}[-4;45]$} & $3^{+37}_{-4}[-6;129]$ &\textcolor{blue}{3$^{+36}_{-4}[-6;124]$} \\
\hline
\end{tabular}

\end{table*}

\subsection{Rotation and dispersion in rotation-dominated disks} \label{ch:3Dfit_results_rot_dom}

The dark shades in Fig.~\ref{fig:violin_Vrot_22rturn}-\ref{fig:violin_Vdisp_mean} show the medians and extents of the kinematical fit parameters of the rotation-dominated systems ($V/\sigma_{\rm V} = 5$ and 10) fitted with either correct or free inclination and/or position angle (the median, scatter, maximum and minimum values of the parameters coloured blue are listed in Table \ref{tab:fit_eval_params}). 

Figure \ref{fig:violin_Vrot_22rturn}-\ref{fig:violin_Vrot_max} and Table \ref{tab:fit_eval_params} show that the extraction radius of the recovered de-projected rotation velocity value is less important than the choice of tool used for the fitting. 
Lower-quality data consistently result in larger ranges of recovered rotation velocity values for all three tools with the strongest effect present for \qubefit. 
\barolo\, consistently overestimates the rotation velocity and exhibits the smallest range of values of the three tools. 
\galpak\, underestimates the rotation velocity and \qubefit\, overestimate it with larger offsets and a significantly larger range of values at lower data quality. 

The velocity dispersion is the most difficult parameter to accurately recover, as seen in Fig.~\ref{fig:violin_Vdisp_22rturn}-\ref{fig:violin_Vdisp_mean}. 
In general, for the three kinematical tools, it is more difficult to recover an accurate dispersion value in higher-quality data, where the output values suffer from larger value ranges, especially strongly so for the parametric tools, and in the case of \galpak\, there are also larger offsets of the median dispersion values in the high-quality data models. 
\barolo\, models show a slightly underestimated dispersion value, and is the tool with the smallest range of dispersion values -- this holds for all data qualities analysed here. 
The parametric tools, \galpak\, and \qubefit, overestimate the recovered median dispersion, with \qubefit's performing better than \galpak. For both tools, the recovered values have large scatters and high maximum offsets, especially on the high-quality data. 
While \galpak\, provides larger median offsets and larger scatters as the data quality increases, \qubefit\, best recovers the medium data quality dispersion values with the smallest value range. 
Furthermore, \qubefit\, shows a clear tendency to better recover the dispersion at lower radii. This is not seen as clearly in the \barolo\, and \galpak\, models, which only show a slight decrease in median offset and value range at lower radii compared to larger radii or the mean dispersion. 

The fourth panels in Fig.~\ref{fig:violin_Vrot_22rturn}-\ref{fig:violin_Vdisp_mean} show the offset in the rotation and dispersion if the values were obtained directly from the synthetically observed cube. 
The dispersion values are worse when obtained directly from the synthetically observed cubes compared to if obtained by any of the 3D-kinematical tools, so this method of value extraction cannot be used for the dispersion values. 
The rotation velocity can, however, be equally well recovered directly from the synthetic observations as via a 3D-kinematical tool for the rotation-dominated systems.

\subsection{Rotation and dispersion in dispersion-dominated systems}\label{ch:3D_fit_param_disp_dom}
We have four intrinsic non-rotation-dominated systems, two fully dispersion dominated, $V/\sigma_{\rm V}=0.2$, one disk with a constant velocity dispersion profile and one with an exponential velocity dispersion profile, and two liminal systems, $V/\sigma_{\rm V}=1$, again one constant and one exponential dispersion profiled system. 
The transparent shading in Fig.~\ref{fig:violin_Vrot_22rturn}-\ref{fig:violin_Vdisp_mean} show the extent and medians of the recovered values for all systems, both the rotation dominated, liminal, and dispersion dominated - displaying the worst case scenario of not knowing if your system is rotation dominated or not. 
However, the \qubefit\, models do not converge often enough for the dispersion-dominated systems to warrant inclusion in this discussion. 
Therefore, we here focus on the \barolo\, and \galpak\, models and note that the \barolo\, models do not converge for low data quality dispersion-dominated systems, and the \galpak\, models do not converge for the high data quality liminal exponential dispersion profiled systems and dispersion-dominated systems.

For the rotation velocity both \barolo\, and \galpak\, suffer from significantly larger ranges of recovered values, although the medians remain stable. 
While the increased ranges follow no clear trend for \barolo, the \galpak\, recovered rotation velocity ranges increase significantly with decreasing data quality, and for low-quality data the recovered value might be 10 times larger than the intrinsic value. 

The velocity dispersion is equally well recovered independent of the $V/\sigma_{\rm V}$ of the input galaxy, apart from an increased range of values from the \barolo\, models -- but this range is still smaller than the \galpak\, recovered range. 

When extracted directly from the synthetically observed cube, the velocity dispersion is still equally poorly recovered, regardless of rotation or dispersion-dominated disks. 
But the inclusion of dispersion-dominated systems cause a significantly increased recovered rotation velocity range. 
Resulting in large uncertainties on both the velocity dispersion and the rotation velocity.

\subsection{The effect of incorrect inclination assumption}\label{ch:3D_fit_section_incl10}
The inclination of a galaxy can be difficult to obtain, so we ran the 3D-fitting tools with two different fixed incorrect inclinations. 
One with the inclination set to 10 degrees above the intrinsic model inclination (inclplus10) and one set to 10 degrees below the intrinsic model inclination (inclmin10). 
The medians and scatter showing the effect of this incorrect inclination assumption on the offsets of the kinematical parameters are marked in Fig.~\ref{fig:violin_Vrot_22rturn}-\ref{fig:violin_Vrot_Vsigma_max_mean_VSintrinsic} in purple and grey. 
The purple lines outline the models fitted at inclination+10 degrees (inclplus10), while the grey outlines the models fitted at inclination-10 degrees (inclmin10). In each case the dotted line includes all disk setups, rotation-dominated, dispersion-dominated, and liminal systems, whereas the solid line outlines the rotation-dominated systems only. 

The most profound effect of an incorrectly assumed inclination is expected to be on the de-projected rotation velocity, as is clear from Eq.~\ref{eq:Vlos_to_Vrot}, which is also reflected in our results. 
Compared to the rotation-dominated systems, assuming a too high inclination angle, inclplus10, results in consistently underestimating the rotation velocity, while not affecting the dispersion. 
On the other hand, assuming a too low inclination angle, inclmin10, leads to consistently overestimating the rotation velocity, and also causes an overestimation of the \galpak\, recovered velocity dispersion.

\subsection{The effect of thick versus thin disk in \galpak}\label{ch:galpak_thick}
\galpak\, offers the option of either fitting the input galaxy with a thick or thin disk together with the choice of profiles for the brightness profile and rotation curve. 
Although our intrinsic disks are modelled as thin disks, thinner than a $0.01''$ pixel thick, we also study the effect of fitting with a thick disk in \galpak. 
The results of the thick disk fits are illustrated in Fig.~\ref{fig:violin_Vrot_22rturn}-\ref{fig:violin_Vdisp_mean} next to the \galpak\, violin plots with black (and grey) squares showing the median values for the rotation-dominated systems (and all systems) and the vertical lines showing the range of recovered values. 
Fitting with a thick disk leads \galpak\, to recover higher rotation velocities than obtained with the thin disk fits, resulting in a more accurate match to the intrinsic value. 
However, as data quality worsens, the range of recovered values increases, eventually exceeding that of the thin disks fits. 
Additionally, thick disk fits yield lower recovered velocity dispersion values, i.e., less overestimated ones than in the thin disk case, and the range of recovered values is also smaller. 
This suggests that, even for intrinsically thin disks, fitting with a thick disk in \galpak\, may provide more accurately recovered rotation velocity and dispersion velocity.

\subsection{Fitting with a constant rotation curve in \qubefit}\label{ch:qubefit_const_rot}
The blue (and cyan) lines in Fig.~\ref{fig:violin_Vrot_22rturn}-\ref{fig:violin_Vdisp_mean} show the median and range, minimum to maximum, of the recovered parameter values for the rotation-dominated systems (all systems). 
The plots show that there is no clear advantage to fitting a galaxy with a constant rotation curve when the data quality is adequate. 
In fact, a higher data quality tends to correlate to a more overestimated recovered velocity dispersion. 
However, when the data quality is poor, using a constant rotation curve when fitting yields a more accurately recovered rotation velocity, without significantly affecting the recovered velocity dispersion value.

\subsection{The choice of velocity dispersion profile with \qubefit}\label{ch:qubefit_expdisp_or_const}
\begin{figure*}
    \centering
    \includegraphics[width=\textwidth]{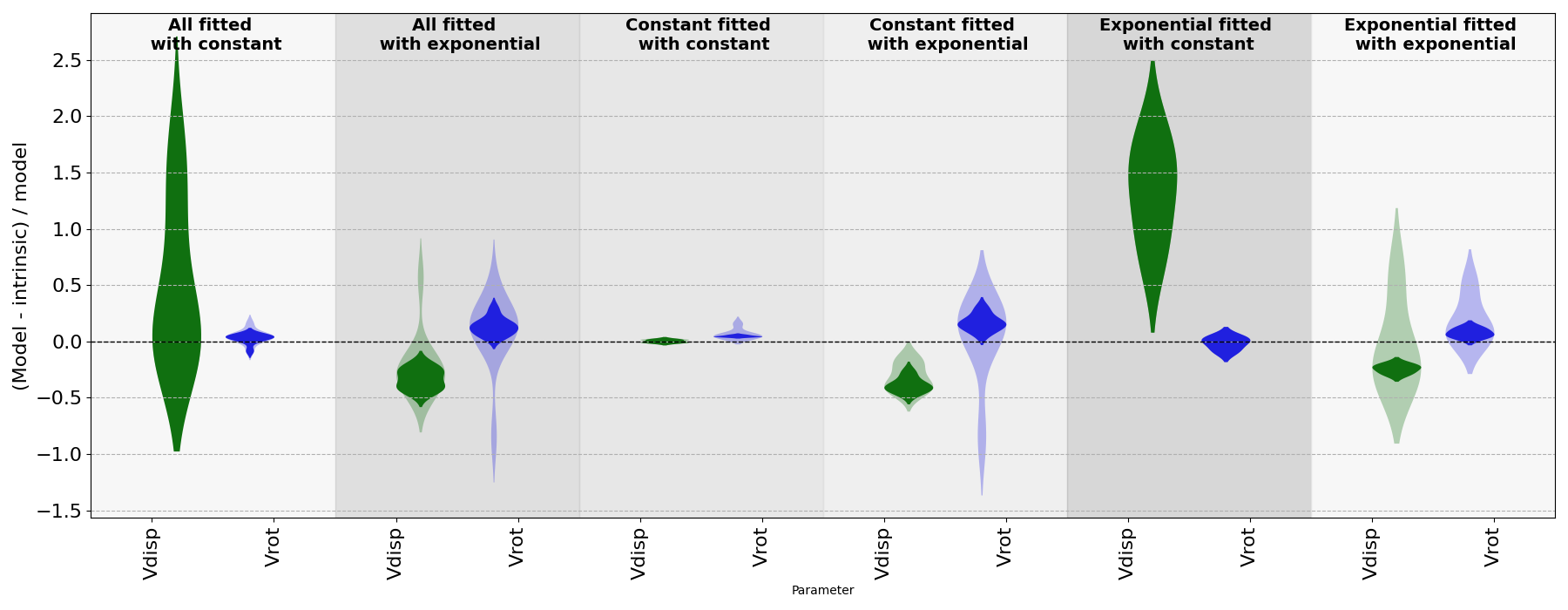}
    \caption{The impact of the choice of velocity dispersion profile when fitting with \qubefit. We show the velocity dispersion and rotation velocity value offsets from the intrinsic at $2.2 r_{\rm d}$ for the high data quality fits, split by intrinsic disks dispersion profiles (constant or exponential) and the corresponding model fits (constant or exponential). The transparent shades indicate the extent of the full data set, including rotation-dominated, liminal, and dispersion-dominated systems, whereas the opaque colours show only the rotation-dominated disks. 
    }
    \label{fig:violin_qubefit_expdispfits}
\end{figure*}

As noted in section \ref{ch:3Dfit_results_rot_dom}, dispersion is the most difficult parameter to accurately recover. 
In \qubefit, we have the option of switching between fitting models with a constant or exponential dispersion profile. 
Figure \ref{fig:violin_qubefit_expdispfits} shows the recovered rotation velocities and dispersions separated by different combinations of intrinsic and model dispersion profiles. 
We only consider values extracted at $2.2 r_{\rm d}$. 
The overall behaviour of the parameter offsets is similar regardless of value extraction method and data quality, though less pronounced at lower data quality. 

The first two sections of Fig.~\ref{fig:violin_qubefit_expdispfits} show the recovered values for all disks, both disks with intrinsic constant dispersion profile and disks with intrinsic exponential dispersion profile. 
It shows that if you do not know whether your input disk has a constant or an exponential dispersion profile, fitting with a constant dispersion profile results in a wider range of dispersion values. 
In contrast, fitting with an exponential dispersion profile instead, on average, leads to an underestimation of the dispersion and an overestimation of the rotation. 
The following four sections split this up into first the disks with constant intrinsic dispersion profile, fitted with constant and exponential dispersion profiles, and the same for the disks with exponential intrinsic dispersion profiles. 
Fitting with an exponential dispersion profile consistently leads to underestimated velocity dispersions and overestimated rotation velocities. 
Additionally, the difference in recovered values between the two fitting methods can be used to infer the intrinsic profile, by first fitting it with a constant and then an exponential dispersion profile. If the change in dispersion value exceeds $150$\% the disk's dispersion most likely follows an exponential profile. 
If the value change is smaller, the intrinsic dispersion shape remains ambiguous. 

Furthermore, a more accurate rotation velocity value can be obtained using a constant dispersion profile fit even when the intrinsic disk dispersion is exponential. 
For large samples where the intrinsic dispersion profile is unknown and fitting resources are limited, an exponential dispersion fit is likely to yield a more concentrated distribution of recovered values, though with an expected underestimation of $\sim 30$\%. 

\section{Discussion} \label{ch:further_discussion}
The sample consists of idealised rotating thin disk models, providing a clean and controlled framework to isolate and investigate the kinematical parameters. 
While these models do not incorporate the full complexity of real galaxies, such as molecular clouds, spiral arms, clumps, outflows, or companions, they serve as an important benchmark. 
For individual galaxies, our results likely represent the best-case scenario. 
However, the trends and conclusions derived from this work are well-suited to reflect the typical behaviours and statistical properties of samples of observed high-redshift disk galaxies.

Furthermore, we note that the thin disk assumption is increasingly accurate for observations of colder gas, as cold gas is likely more concentrated along the galaxy mid plane. Ionised gas, such as H$\alpha$, may require caution as ionised gas is more sensitive to outflowing gas components and H$\alpha$ has been seen to reside in a thicker disk formation \citep{kohandel2024}. 
Gaseous non-planar structures are interesting as they are located above and below the galaxy mid plane and rotate slower. 
Fortunately, the bulk of gaseous emission is in galaxy mid planes, which for these high-redshift synthetic observations result in a further decreased concern for any non-mid plane emission. 
Furthermore, in Chapter \ref{ch:galpak_thick}, we noted that fitting high-quality observations of thin disks with a thick disk fit setting in \galpak, can more accurately recover rotation and dispersion velocities, indicating that adopting a thick disk fit setting in \galpak may be the better choice. 

We have noted that the results presented in this paper can be extrapolated to any emission line observed in a high-$z$ galaxy with ALMA, {\it JWST}, or similar observatories, as long as the emission traces the galaxy disk. 
However, we further note that {\it JWST} observations of high-$z$ galaxies is primarily of ionised gas, such as H$\alpha$, which is expected to show higher velocity dispersions than the colder gas tracers. This may result in disks that appear more dispersion dominated, and as we have seen, in Table \ref{tab:fit_eval_params} and Chapter \ref{ch:3D_fit_param_disp_dom}, systems that are more dispersion dominated result in a lower convergence rate for the 3D-kinematical tools and a significantly increased range of recovered rotation velocity values. This indicates that individual source studies comes with even further risks, highlighting that averages of large samples of galaxies is the more reliable method. 
Furthermore, though {\it JWST} NIRSpec and MIRI IFU's boast high spatial resolution, similar to our high and medium data quality, the spectral resolution suffers, ranging from 30 to 230 km/s, significantly larger than our ALMA 10 km/s velocity bins. This likely complicates the recovery of the dispersion velocity, although as the dispersion velocity is expected to be higher for ionised gas compared to the work presented in this paper, this may cancel out. 
It would be beneficial to carry out similar studies to this one on synthetic ionised gas integral field spectroscopy observations to further detail the impact of the chosen gas tracer and the instrument specific PSF on the recovered rotation velocity and velocity dispersion. 


\subsection{On the 3D-fitting tools} \label{ch:3Dfit_discussion}
At the outset of this project, we hypothesised that the choice of flux, rotation velocity, and velocity dispersion profiles in the 3D-fit settings of the parametric tools, \galpak\, and \qubefit, would have a significant impact on the outcome of the fits. 
This enforced choice and restriction of the user predicting the moment profiles prior to the fitting could be expected to bias the parametric tools to a higher degree than a tool relying on tilted ring models, such as \barolo. 
In addition, the higher number of free parameters in the \barolo\, fits compared to the parametric tools, should result in a model with a lower residual, albeit with an increased risk of overfitting the data. 
We see a larger range of recovered velocity dispersion values for the parametric tools compared to the tilted ring method. 
Figure \ref{fig:violin_Vrot_22rturn}-\ref{fig:violin_Vdisp_mean} show that, considering the offsets and value ranges of the recovered rotation velocity and dispersion values, \barolo\, is in the lead. 
However, the information that a tilted ring model and its residuals can reveal regarding the physical properties of a galaxy is significantly harder to interpret, and individual rings can fully trace non-circular motions instead of the disk, depending on the relative emission strength. 
Fortunately, bars, and some outflows, extend only in the central regions of the kinematical fields, allowing a tilted ring model to adjust for the inner non-axisymmetries with the innermost rings and fit the disk well in the outer regions, while a parametric model is likely biased by a bar or outflow though the entire disk model. 
The parametric models have the advantage of more clearly separating what part of a source can be seen as a rotating disk and what part is non-axisymmetric, simplifying the interpretation of the residuals and highlighting non-axisymmetric structures -- especially if an outflow can be separated and blocked out prior to fitting. 

Regarding the choice of moment profiles in the parametric models, we support \citet{lee2024}'s conclusion that the profiles have to be chosen with care. 
\citet{lee2024}'s galaxies are at $z=1-3$ with instrumental effects added as random Gaussian noise and beam convolution, their models are created using the respective 3D-kinematical fitting tools, and the geometrical parameters are set to the intrinsic values, whereas our disks are at $z=5$, created independently from the respective 3D-kinematical tools, are synthetically observed with ALMA, and we allow for variation of position angles and inclinations. 
\citet{lee2024} find a strong dependence of the ability to recover the kinematical parameters on the exact form of the input galaxy, the combination of moment major axis profiles/matter distribution. We also note that, in our galaxies the largest scatter is caused by the unknown input intrinsic galaxy profile combinations, whereas the choice of profiles for the fitting with parametric tools has a smaller effect on the scatter of recovered parameter values. 

Focusing on the choice of velocity dispersion profile, we see in Fig.~\ref{fig:violin_qubefit_expdispfits} that depending on the choice we can either end up overestimating the rotation velocity while underestimating the dispersion -- resulting in overestimating the abundance of cold disks (fitting with an exponential dispersion profile) -- or we can overestimate the dispersion and thereby underestimate the abundance of cold disks (fitting with a constant dispersion on a disk with an exponential dispersion). 
It is also interesting to note that \galpak\, suffers from particular difficulties in converging when fitting systems with exponential velocity dispersion profiles (as seen in Table \ref{table:3Dconvergence_galpak}). 

Furthermore, we note that using a constant rotation curve for fitting the disks is advantageous in low-quality data, and may there result in more accurately recovered velocities. 
As may fitting with a thick disk in \galpak, even though the input disk is known to be thin, but more investigation is required for a conclusion on the topic.


\subsection{The presence of dynamically cold rotating disks and $V/\sigma_{\rm V}$}

\begin{figure*}
    \centering
    \includegraphics[width=\textwidth]{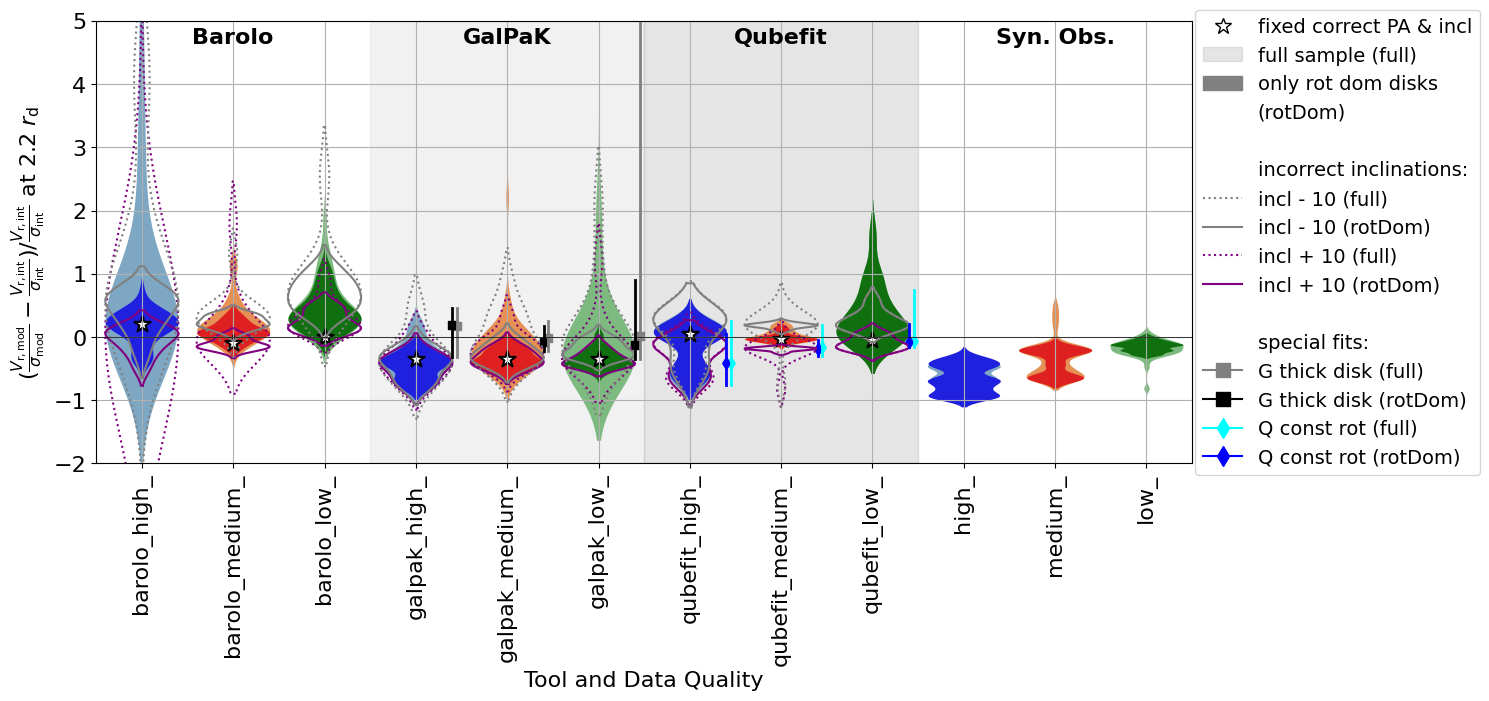}
    \caption{The recovery of $V_{\text{rot}}/\sigma_{\rm V}$ at $2.2 r_{\rm d}$. The figure is colour-coded as Fig.~\ref{fig:violin_Vrot_22rturn}.}
    \label{fig:violin_Vrot_Vsigma_22rturn_VSintrinsic}
\end{figure*}

\begin{figure*}
    \centering
    \includegraphics[width=\textwidth]{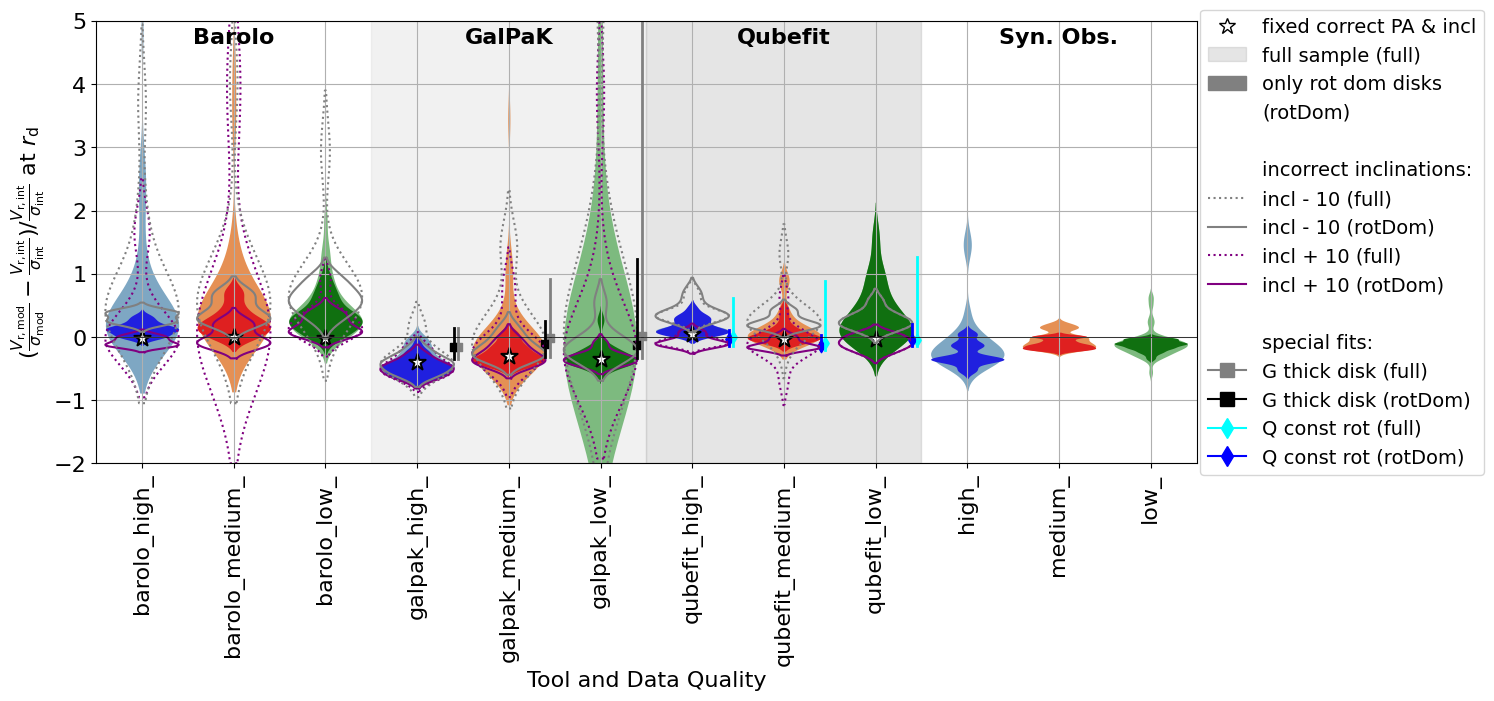}
    \caption{The recovery of $V_{\text{rot}}/\sigma_{\rm V}$ at $1.0 r_{\rm d}$. The figure is colour-coded as Fig.~\ref{fig:violin_Vrot_22rturn}.}
    \label{fig:violin_Vrot_Vsigma_rturn_VSintrinsic}
\end{figure*}

\begin{figure*}
    \centering
    \includegraphics[width=\textwidth]{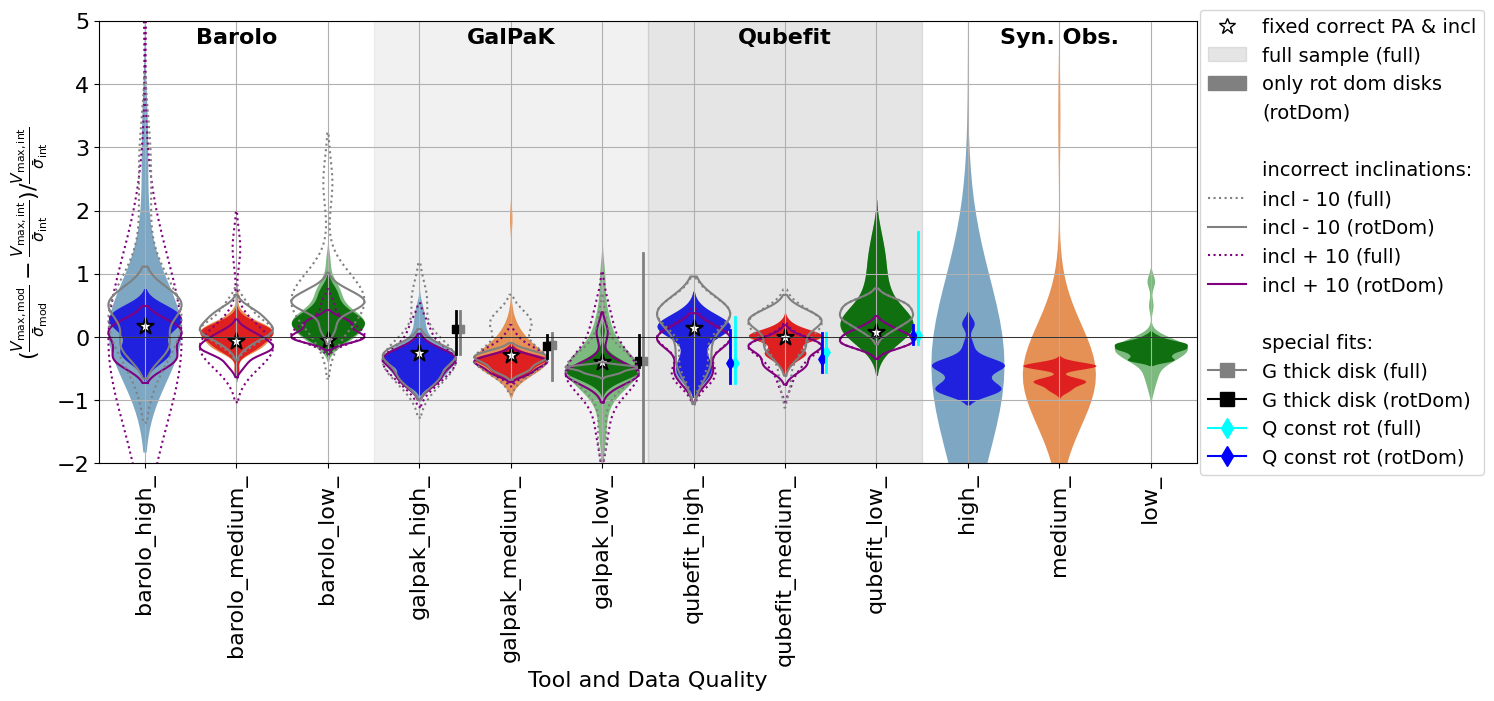}
    \caption{The recovery of $V_{\text{rot, max}}/\bar{\sigma_{\rm V}}$, where $V_{\text{rot, max}}$ is the maximum de-projected rotation velocity within the maximum disk radius and $\bar{\sigma_{\rm V}}$ the average dispersion within the same radius. The figure is colour-coded as Fig.~\ref{fig:violin_Vrot_22rturn}. }
    \label{fig:violin_Vrot_Vsigma_max_mean_VSintrinsic}
\end{figure*}

\setlength{\tabcolsep}{1pt}
\begin{table*}
\scriptsize
    \caption{
    The deviation of $V/\sigma_{\rm V}$ from the intrinsic values as recovered by \barolo, \galpak, and \qubefit converged models. We present the median, $16^{\rm th}$, and $84^{\rm th}$ percentile values of the $V/\sigma_{\rm V}$ distribution in percentages, plus the minimum and maximum percentage values inside brackets (Median$^{rel.84\%}_{rel.16\%}[min;max]$). 
    ``(All)'' denotes the sample containing both dispersion-dominated, liminal, and rotation-dominated disks, whereas the right-hand values, ``(RotDom)'', in blue are for the rotation-dominated disks ($V/\sigma_{\rm V} > 1$) only.}
    \label{tab:VrotVdisp_simalmaVSintrinsic_comparison}
    \centering
    \begin{tabular}{lcccccccc} \hline  \hline 
          & \multicolumn{2}{c}{High data quality} & \multicolumn{2}{c}{Medium data quality} & \multicolumn{2}{c}{Low data quality} & \multicolumn{2}{c}{All data qualities} \\
          Parameter & (All) & {\color{blue}(Rot Dom)} & (All) & {\color{blue}(Rot Dom)} & (All) & {\color{blue}(Rot Dom)} & (All) & {\color{blue}(Rot Dom)} \\ \hline 
    \noalign{\smallskip}
    \textbf{Synthetic obs.}  &    &       &        &       &      &      &      &  \\    \smallskip 
         $V/\sigma_{\rm V}$ at $2.2r_{\rm d}$ & $-65^{+21}_{-27}[-96;-29]$ &\textcolor{blue}{$-67^{+23}_{-26}[-96;-29]$} & $-24^{+6}_{-40}[-67;48]$ &\textcolor{blue}{$-33^{+13}_{-32}[-67;-16]$} & $-18^{+12}_{-10}[-81;7]$ &\textcolor{blue}{$-14^{+7}_{-10}[-28;1]$} & $-34^{+20}_{-32}[-96;48]$ &\textcolor{blue}{$-28^{+15}_{-39}[-96;1]$} \\ \smallskip 
        $V/\sigma_{\rm V}$ at $r_{\rm d}$ & $-19^{+28}_{-18}[-54;164]$ &\textcolor{blue}{$-33^{+22}_{-5}[-54;7]$} & $-7^{+21}_{-11}[-22;22]$ &\textcolor{blue}{$-12^{+14}_{-6}[-22;2]$} & $-13^{+11}_{-15}[-56;64]$ &\textcolor{blue}{$-11^{+8}_{-13}[-34;-1]$} & $-13^{+21}_{-17}[-56;164]$ &\textcolor{blue}{$-14^{+11}_{-20}[-54;7]$} \\ \smallskip 
        $V_\text{rot,max}/\bar{\sigma_{\rm V}}$ & $-42^{+40}_{-36}[-89;1746]$ &\textcolor{blue}{-50$^{+19}_{-34}[-89;22]$} & $-48^{+6}_{-33}[-86;1000]$ &\textcolor{blue}{-57$^{+13}_{-21}[-86;-40]$} & $-19^{+15}_{-18}[-84;88]$ &\textcolor{blue}{-17$^{+10}_{-12}[-37;3]$} & $-42^{+35}_{-33}[-89;1746]$ &\textcolor{blue}{-43$^{+32}_{-34}[-89;22]$} \\ 
        \noalign{\smallskip}
              \textbf{\barolo}&    &       &        &       &      &      &      &  \\ \smallskip 
        $V/\sigma_{\rm V}$ at $2.2r_{\rm d}$ & $25^{+135}_{-27}[-38;916]$ &\textcolor{blue}{21$^{+22}_{-29}[-38;71]$} & $13^{+27}_{-17}[-21;117]$ &\textcolor{blue}{7$^{+16}_{-12}[-21;40]$} & $35^{+59}_{-22}[0;185]$ &\textcolor{blue}{35$^{+33}_{-23}[0;115]$} & $25^{+57}_{-21}[-38;916]$ &\textcolor{blue}{20$^{+23}_{-22}[-38;115]$} \\ \smallskip 
        $V/\sigma_{\rm V}$ at r$_{\rm d}$ & $19^{+71}_{-15}[-3;482]$ &\textcolor{blue}{12$^{+11}_{-11}[-3;34]$} & $29^{+52}_{-23}[-1;435]$ &\textcolor{blue}{21$^{+28}_{-18}[-1;72]$} & $32^{+46}_{-24}[-1;185]$ &\textcolor{blue}{29$^{+30}_{-22}[-1;102]$} & $27^{+56}_{-21}[-3;482]$ &\textcolor{blue}{19$^{+23}_{-16}[-3;102]$} \\ \smallskip 
        $V_\text{rot,max}/\bar{\sigma_{\rm V}}$ & $18^{+40}_{-42}[-41;909]$ &\textcolor{blue}{16$^{+23}_{-43}[-41;48]$} & $5^{+15}_{-20}[-52;67]$ &\textcolor{blue}{5$^{+13}_{-20}[-52;32]$} & $27^{+40}_{-26}[-10;122]$ &\textcolor{blue}{24$^{+26}_{-23}[-7;82]$} & $16^{+29}_{-30}[-52;909]$ &\textcolor{blue}{15$^{+23}_{-29}[-52;82]$} \\
         \noalign{\smallskip}
             \textbf{\galpak}&    &       &        &       &      &      &      &  \\ \smallskip 
         $V/\sigma_{\rm V}$ at $2.2r_{\rm d}$ & $-40^{+22}_{-28}[-78;20]$ &\textcolor{blue}{$-40^{+8}_{-29}[-78;-17]$} & $-23^{+29}_{-21}[-56;221]$ &\textcolor{blue}{$-37^{+17}_{-12}[-56;-6]$} & $-24^{+50}_{-19}[-57;674]$ &\textcolor{blue}{-36$^{+16}_{-11}[-57;4]$} & $-34^{+41}_{-14}[-78;674]$ &\textcolor{blue}{$-38^{+18}_{-15}[-78;4]$} \\ \smallskip 
        $V/\sigma_{\rm V}$ at r$_{\rm d}$ & $-44^{+23}_{-17}[-64;-1]$ &\textcolor{blue}{$-46^{+19}_{-15}[-64;-8]$} & $-17^{+28}_{-23}[-46;345]$ &\textcolor{blue}{$-32^{+23}_{-9}[-46;7]$} & $-22^{+97}_{-18}[-55;1551]$ &\textcolor{blue}{$-35^{+14}_{-7}[-55;31]$} & $-27^{+39}_{-18}[-64;1551]$ &\textcolor{blue}{$-35^{+18}_{-11}[-64;31]$} \\ \smallskip 
        $V_\text{rot,max}/\bar{\sigma_{\rm V}}$  & $-34^{+19}_{-27}[-72;45]$ &\textcolor{blue}{$-35^{+12}_{-27}[-72;-11]$} & $-32^{+26}_{-16}[-59;188]$ &\textcolor{blue}{$-39^{+9}_{-15}[-59;-20]$} & $-51^{+43}_{-21}[-251;94]$ &\textcolor{blue}{$-52^{+16}_{-13}[-80;10]$} & $-37^{+29}_{-23}[-251;188]$ &\textcolor{blue}{$-44^{+14}_{-16}[-80;10]$} \\

         \noalign{\smallskip}
              \textbf{Qubefit}&    &       &        &       &      &      &      &  \\ \smallskip 
         $V/\sigma_{\rm V}$ at $2.2r_{\rm d}$ & $5^{+10}_{-69}[-68;28]$ &\textcolor{blue}{5$^{+6}_{-63}[-67;28]$} & $-1^{+15}_{-4}[-11;21]$ &\textcolor{blue}{-2$^{+8}_{-5}[-11;16]$} & $21^{+67}_{-25}[-11;171]$ &\textcolor{blue}{21$^{+67}_{-25}[-11;171]$} & $5^{+20}_{-13}[-68;171]$ &\textcolor{blue}{2$^{+26}_{-9}[-67;171]$} \\\smallskip 
        $V/\sigma_{\rm V}$ at r$_{\rm d}$ & $12^{+18}_{-5}[3;46]$ &\textcolor{blue}{12$^{+20}_{-8}[3;46]$} & $4^{+29}_{-8}[-8;91]$ &\textcolor{blue}{1$^{+17}_{-5}[-8;46]$} & $19^{+67}_{-22}[-15;167]$ &\textcolor{blue}{19$^{+67}_{-22}[-15;167]$} & $11^{+33}_{-13}[-15;167]$ &\textcolor{blue}{10$^{+27}_{-13}[-15;167]$} \\ \smallskip 
        $V_\text{rot,max}/\bar{\sigma_{\rm V}}$ & $13^{+6}_{-72}[-62;39]$ &\textcolor{blue}{13$^{+7}_{-63}[-62;39]$} & $-2^{+6}_{-24}[-44;20]$ &\textcolor{blue}{0$^{+5}_{-28}[-44;20]$} & $27^{+72}_{-19}[-13;169]$ &\textcolor{blue}{27$^{+72}_{-19}[-13;169]$} & $6^{+23}_{-34}[-62;169]$ &\textcolor{blue}{9$^{+25}_{-37}[-62;169]$} \\
        \noalign{\smallskip}
         \hline 
    \end{tabular} \\
\end{table*}

The discussion of the presence and abundance of cold rotating disks in the early Universe is of particular importance for our understanding of galaxy evolution and its history. 
Assessing the ''coldness'', i.e., the rotational dominance in a disk is done via the ratio between the rotation velocity and velocity dispersion, $V/\sigma_{\rm V}$. 
Caution is advised as these parameters can be obtained and extracted in a multitude of ways. 
In this work, we present the $V/\sigma_{\rm V}$ ratio derived from the values recovered by the 3D-kinematical tools (Table \ref{tab:fit_eval_params}), as well as the values extracted directly from the synthetically observed cubes.

\subsubsection{3D-modelled $V/\sigma_{\rm V}$ in rotation-dominated disks}
The median offset, $1\sigma$ scatter, and maximum and minimum values of the rotation velocity and velocity dispersion values for the three data qualities are listed in Table \ref{tab:VrotVdisp_simalmaVSintrinsic_comparison}. 
The blue coloured numbers note the rotation-dominated systems, and plotted in Fig.~\ref{fig:violin_Vrot_Vsigma_22rturn_VSintrinsic}-\ref{fig:violin_Vrot_Vsigma_max_mean_VSintrinsic}. 

The $V/\sigma_{\rm V}$ ratio portrays a pretty consistent median offset and scatter per 3D-kinematical tool regardless of method of derivation, despite separately large value ranges for the rotation velocity and dispersion velocity. 
The dispersion is the most difficult parameter to recover for the 3D-kinematical tools, as \citet{lee2024} also note, and causes the largest scatter of recovered values for high-quality data. 
This may be in part due to the differences in how the three tools model dispersions, and the largest effect, which is seen in \galpak, may be mitigated by using a thick disk fit, despite our disks being intrinsically modelled as thin. 
The dispersion is more accurately reproduced by the 3D models at lower radial distances, which of course leads to a limited view of the disk kinematics. 
The dispersion is also more accurately recovered in lower-quality data, likely due to the smoothing out of variations across the disks. 
We also further note that the intrinsic versus 3D-modelled dispersion profile choice has a strong effect on the recovered dispersion value, see Fig.~\ref{fig:violin_qubefit_expdispfits}. 
But the reason for the large scatter and value range of the velocity dispersion remains unclear. 
The rotation velocity is, on the other hand, more accurately recovered, in high and medium-quality data. 
The recovered rotation velocity shows a significantly larger range of values at low data quality, especially for the parametric tools. 
As the recovered dispersion velocity follows an opposite trend with small range for low data quality and a large range for high data quality, the resultant $V/\sigma_{\rm V}$ ratio is at an advantage. 

Although, the low-quality data are not a recommended data quality to use for resolved kinematics studies, the median offsets for the $V/\sigma_{\rm V}$ values are comparable to the high and medium quality data and may therefore be useful for large samples of rotation-dominated systems, where \galpak\, in particular provides the smallest range of output parameters. 

Focusing on the medium and high data quality systems, \barolo, on average, overestimates the rotation velocity and underestimates the velocity dispersion, resulting in an overestimation of the abundance of cold disks by $V/\sigma_{\rm V} = [(+5^{13}_{-20})-(+21^{28}_{-18})]\%$ (depending on method of value extraction) for samples of rotation-dominated systems. 
The median offset is within the $1\sigma$ uncertainties, implying that the abundance of cold disks is adequately recovered. 
However, for single sources, the $V/\sigma_{\rm V}$ can be as high as +72\% larger or $-$52\% smaller than the actual value. 

\galpak\, tends toward underestimation of the rotation velocity and overestimation of the dispersion, resulting in the largest underestimation of $V/\sigma_{\rm V}$. 
On average for rotation-dominated systems the abundance and presence of cold disks will be underestimated by on average $(-32 ^{+23}_{-9})-(-52^{+16}_{-13})$\% in high and medium-quality data depending on method of derivation. For single sources the range of possible values is $-$78\% to +7\%. 

\qubefit\, recovers the most accurate median $V/\sigma_{\rm V}$ in high and medium data quality rotation-dominated systems, and performs worst at low data quality. 
For the medium and high-quality data the $V/\sigma_{\rm V}$ ranges $(-2^{+8}_{-5}) - (13^{+7}_{-63})$\%, with a maximum and minimum of +46\% and $-$67\% respectively.

\subsubsection{3D-modelled $V/\sigma_{\rm V}$ with incorrect inclination}
From Fig.~\ref{fig:violin_Vrot_Vsigma_22rturn_VSintrinsic}-\ref{fig:violin_Vrot_Vsigma_max_mean_VSintrinsic} we see the effect of incorrectly assumed inclinations. 
As expected, a too high assumed inclination causes an underestimated $V/\sigma_{\rm V}$ and a too low assumed inclination causes an overestimated $V/\sigma_{\rm V}$ in \barolo\, and \qubefit. 
This is the expected behaviour due to the effect of inclination on the de-projected rotation velocity. 
However, we do not see this effect on the \galpak\, recovered $V/\sigma_{\rm V}$, as the \galpak\, models result in less of an effect on the recovered rotation velocity the assumed inclination is too high, and for the too low assumed inclination \galpak\, overestimates the dispersion, which balances out the recovered $V/\sigma_{\rm V}$ value.

\subsubsection{3D-modelled $V/\sigma_{\rm V}$ in non-optimal disks}
When we increase the dispersion relative to the rotation velocity the ability of the 3D-kinematical tools to converge is severely hampered (see Table \ref{table:3Dconvergence_barolo}-\ref{table:3Dconvergence_qubefit}). The \barolo\, convergence rate declines from 100\% to 81\%, \galpak\, from 92\% to 73\%, and \qubefit\, all the way from 99\% to 46\% -- too low to enable  
conclusions to be drawn. 
The transparent shades in Fig.~\ref{fig:violin_Vrot_Vsigma_22rturn_VSintrinsic}-\ref{fig:violin_Vrot_Vsigma_max_mean_VSintrinsic} together with black-coloured numbers in Table \ref{tab:VrotVdisp_simalmaVSintrinsic_comparison} show how large the scatter and value range become for \barolo\, and \galpak\, when the input systems are not solely rotation dominated. 
These values are indicative of what to expect when the input system is unknown. 

For high-quality data, \galpak\, is likely to provide a comparably good fit as it does for the pure rotation-dominated disks, with a $V/\sigma_{\rm V}$ underestimation of $-40^{+22}_{-78}$\% (min; max = -78; 20\%) (depending on method of value extraction). 
But for low-quality data the value range can be as large as -251\% to +1551\%. 

\barolo\, on the other hand, handles low-quality data better while suffering with high-quality data when the input disks are not purely rotation dominated. 
The median offsets are similar to those of purely rotation dominated samples, with only slightly higher overestimations of $V/\sigma_{\rm V}$, at $(+18^{+40}_{-42})-(+35^{+59}_{-22})\%$ but ranging from -41\% to +916\% at high data quality and -10\% to 185\% at low data quality. 

However, the large value ranges of $V/\sigma_{\rm V}$ propagated by the dispersion-dominated and liminal systems are presented as percentage offsets from the intrinsic values. And for these systems the intrinsic value is 0.2 and 1, so while this can push the system into the region of dynamically cold disks, the likelihood of that happening is not as pronounced as the percentages make it appear. 
Still, considering the value ranges there is still a significant risk of overestimating the $V/\sigma_{\rm V}$ and thereby overestimating how dynamically cold the observed system is. 
Fortunately, this mixed rotation and dispersion dominated sample still shows stable median $V/\sigma_{\rm V}$ offsets, so for large samples of galaxies of unknown $V/\sigma_{\rm V}$ range we may still obtain adequate $V/\sigma_{\rm V}$ values. 
But to achieve a kinematically accurate model, it is vital to know whether the input disk is rotation dominated or not.

\subsubsection{$V/\sigma_{\rm V}$ from synthetic observations instead of 3D-fitted models}
Throughout this paper we have plotted the rotation velocity, velocity dispersion, and $V/\sigma_{\rm V}$ extracted from \barolo, \galpak, \qubefit\, models, as well as extracted directly from the synthetically observed cube. 
We have noted that the rotation velocities extracted directly from the synthetically observed cubes suffer from large scatter when the input systems are not purely rotation dominated, worse for higher data quality.
Similarly, we find that the extracted velocity dispersions are more inaccurate and scattered at higher data quality. 
But again, the $V/\sigma_{\rm V}$ ratio partly compensates for these discrepancies, and Fig.~\ref{fig:violin_Vrot_Vsigma_22rturn_VSintrinsic}-\ref{fig:violin_Vrot_Vsigma_rturn_VSintrinsic} and Table \ref{tab:VrotVdisp_simalmaVSintrinsic_comparison} show that if the $V/\sigma_{\rm V}$ value is obtained from a rotation velocity and dispersion extracted at the same radial distance along the major axis, the resultant $V/\sigma_{\rm V}$ value is good; 
The median offset, scatter and value range obtained for a mixed rotation- and dispersion-dominated sample are smaller when extracted directly from the synthetically observed cube at $2.2 r_{\rm d}$ ($V/\sigma_{\rm V} = -34^{+20}_{-32}\%$, min;max=-96;48\%) and $1r_{\rm d}$ ($V/\sigma_{\rm V} = -13^{+21}_{-17}\%$, min;max=-56;164\%) than if obtained from the 3D models (independent of data quality). 
This is particularly important when there is a risk of dispersion-dominated systems, where the scatter and value range are significantly reduced when extracted directly from the synthetically observed cube. 

However, for the $V_{\rm rot,max}/\bar{\sigma_{\rm V}}$ the median offset, scatter and value range are larger and therefore not reliable. 
Using this method of $V/\sigma_{\rm V}$ derivation on the observed cube is therefore not recommended. 

In conclusion, when including the dispersion-dominated systems, fitting the data with a 3D-kinematical tool does not provide better $V/\sigma_{\rm V}$ values than directly extracting the values from the observed cube as long as the values are extracted at specific radii along the major axis ($r_{\rm d}$ and $2.2 r_{\rm d}$). 
For the rotation dominated input galaxies, direct extraction $V/\sigma_{\rm V}$ at $r_{\rm d}$ and $2.2 r_{\rm d}$, and $V/\sigma_{\rm V}$ as $V_{\rm rot,max}/\bar{\sigma_{\rm V}}$, produces values comparable to those derived from the 3D-fitted cubes. 
Extraction of $V/\sigma_{\rm V}$ directly from the observed cube may therefore be used for classifying an observed galaxy as rotation dominated enough to be fitted with a 3D-kinematical tool or not. 
If the observed data is poor the best (closest median value and smallest scatter) $V/\sigma_{\rm V}$ values can be obtained from direct extraction along the major axis of the observed cube. 
This is under the caveat that the system consists of a single disk with a combination of moment profiles and rotation velocity to velocity dispersion ratios within the range investigated by our setup. 
Machine learning classification algorithms  and/or methods to narrow down the geometrical parameters of galaxies (e.g., CANNUBI \citep{romanoliveira2023}) may be crucial for ensuring individual disk and rotation domination in a system prior to 3D-fitting, to facilitate trustworthy results.

\section{Summary and conclusions}
\label{ch:sum}
We investigated the uncertainties and systematics of three 3D kinematic tools, \barolo, \galpak, and \qubefit, in recovering the intrinsic kinematics of rotating disks at high-redshift. 
We built a large sample of idealized galaxy disk models. 
These disks were synthetically observed with ALMA, creating realistic interferometric synthetic data corresponding to line emission of the \cii 1900.536\,GHz emission line at a redshift of 5.1 (although the results presented here can be applied to any emission line that traces the galactic disk).

\smallskip
\noindent 
We summarise the key results as follows: 
\begin{itemize}    
    \item  The recovery of the rotation velocity and velocity dispersion is dependent on the choice of 3D-kinematical tool. For samples of rotation-dominated systems with kinematic parameters extracted at $2.2r_{\rm d}$ and averaged over data quality: 
     
    \begin{itemize}

        \item \barolo\, overestimates the rotation velocity and underestimates the velocity dispersion, resulting in an overestimation of the abundance of cold rotating disks, by $+20^{+23}_{-22}\%$.  

        \item \galpak, on the other hand, underestimates the rotation velocity and overestimates the dispersion, thereby underestimating the $V/\sigma_{\rm V} = -38^{+18}_{-15}\%$. 

        \item \qubefit\, overestimates both the rotation velocity and dispersion, which results in a better $V/\sigma_{\rm V}$ of $2^{+26}_{-9}\%$. 

        \item All three tools provide qualitatively similar $V/\sigma_{\rm V}$ ratios for all three extraction methods explored (i.e., extracting at $2.2r_{\rm d}$, $r_{\rm d}$, and as $V_{\rm rot,max}/\bar{\sigma_{\rm V}}$). 

        \item For individual sources the range of recovered values can be large. The rotation velocity value range increases as data quality decreases. The velocity dispersion is the most difficult to constrain for the parametric tools, \galpak\, and \qubefit, in high-quality data. \barolo\, provides the smallest range of recovered dispersion values, in general, which indicates that \barolo\, might be the most suitable for obtaining accurate dispersion velocities (although underestimated) for individual sources. 
        
    \end{itemize}

    \item Normally, we do not know if our sample sources are rotation-dominated prior to fitting. 
    \qubefit\, rarely converges for dispersion-dominated systems, but \barolo\, and \galpak\, exhibit a higher prevalence of convergence. For samples with a mix of rotation-dominated ($V/\sigma_{\rm V}=10, 5$), liminal ($V/\sigma_{\rm V}=1$), and dispersion-dominated systems ($V/\sigma_{\rm V}=0.2$): 

    \begin{itemize}
    
        \item Decreasing data quality correlates to an increasing range of recovered rotation velocity values, for both \barolo\, and \galpak, with the strongest effect seen in the \galpak\, fits. \\
        In general fitting the non-rotation-dominated systems result in a significantly increased scatter on all recovered parameter values. However, the averages remain comparable to the median offsets obtained for rotation dominated samples, so while individual fitted values cannot be trusted, the averages of large samples are still likely to provide useful characterisation of the sample as a whole.

    \end{itemize}

    \item For samples of rotation-dominated disks, extracting the $V/\sigma_{\rm V}$ directly from the synthetically observed cubes provides slightly worse but still comparable median offsets and scatters as the values extracted from the 3D-modelled cubes if the rotation velocity and dispersion values are both extracted at the same radius. However, for mixed samples, containing dispersion dominated and liminal systems as well, direct extraction from the observed cubes results in similar median offsets but with smaller value ranges, especially so for lower quality data. 
    Therefore it may be advantageous to obtain the $V/\sigma_{\rm V}$ directly from the observed cube to indicate whether the target is rotation dominated enough to be fitted with a 3D-kinematical tool. 
    However, note that this is only applicable if the rotation velocity and dispersion are extracted at the same radius.

    \item Although low-quality data (2-3 beams across the major axis) is not recommended for resolved galaxy kinematics the recovered median values of the rotation velocity and velocity dispersion for samples remain adequate, indicating that fitting samples of low resolution data can still provide usable average values - with the median offsets presented in this paper taken into account.

    \item In the parametric tools, \galpak\, and \qubefit, the combination of brightness profile, rotation curve, and dispersion profile has to be set prior to fitting. Using a constant rotation curve, an option in \qubefit, when fitting is advantageous for poorly resolved data, and, surprisingly, not detrimental to higher resolution data either. 
    With \galpak, we see indications that fitting intrinsically thin disks with thick disk fits may result in smaller kinematical parameter value offsets. 
    The choice of dispersion profile (constant versus exponential, option in \qubefit) has a direct impact on the recovered rotation velocity and velocity dispersion, and requires careful handling and assessment during the fitting procedure.

\end{itemize}

The derived abundance and presence of dynamically cold disks in a galaxy sample, as quantified by the $V/\sigma_{\rm V}$ ratio, depend on the choice of 3D-kinematical fitting tool. 
For the same sample, different tools may lead to either an overestimation or underestimation of the prevalence of cold disks. 
These findings are especially relevant in the context of recent high-redshift observations with ALMA and JWST, where kinematical classifications play a key role in tracing the evolutionary pathways of galaxies.
Tool-dependent biases in recovered $V/\sigma_{\rm V}$ values must therefore be carefully considered to avoid systematic misclassification of galaxies. 

We recommend that any kinematic parameter derived from 3D modelling be interpreted in light of the median offsets and scatters presented in this work. 
Applying correction factors or bias estimates, such as those provided here, and deliberately selecting the 3D-fitting tool will aid in drawing robust scientific conclusions.
The range of recovered kinematic parameter values can be substantial for individual sources. 
This study offers insight into how far one can trust fitted values for individual sources, and provides expectations for median offsets when interpreting samples.

\section*{Acknowledgements}
  M.Y.\ and K.K.\ acknowledges support via research grants from the Knut and Alice Wallenberg Foundation (KAW 2017.0292 and 2019.0443). J.M.\ gratefully acknowledge financial support from ANID - MILENIO - NCN2024\_112. 
  G.C.J. acknowledge support by the Science and Technology Facilities Council (STFC), by the ERC through Advanced Grant 695671 ''QUENCH'', and by the UKRI Frontier Research grant ''RISEandFALL''. 
  M.Y. expresses appreciation to Dr. Marcel Neeleman, Dr. Nicolas Bouch\'e, and Dr. Fernanda Roman de Oliveira for their availability and willingness to answer questions and discuss \qubefit, \galpak, and \barolo\, respectively. 
  We thank the referee for their useful comments and input that improved the manuscript.

\section*{Data Availability}
The data underlying this article will be shared on reasonable request to the corresponding author.


\bibliographystyle{mnras}
\bibliography{bibliography}

\begin{thebibliography}{}
\makeatletter
\relax
\def\mn@urlcharsother{\let\do\@makeother \do\$\do\&\do\#\do\^\do\_\do\%\do\~}
\def\mn@doi{\begingroup\mn@urlcharsother \@ifnextchar [ {\mn@doi@}
  {\mn@doi@[]}}
\def\mn@doi@[#1]#2{\def\@tempa{#1}\ifx\@tempa\@empty \href
  {http://dx.doi.org/#2} {doi:#2}\else \href {http://dx.doi.org/#2} {#1}\fi
  \endgroup}
\def\mn@eprint#1#2{\mn@eprint@#1:#2::\@nil}
\def\mn@eprint@arXiv#1{\href {http://arxiv.org/abs/#1} {{\tt arXiv:#1}}}
\def\mn@eprint@dblp#1{\href {http://dblp.uni-trier.de/rec/bibtex/#1.xml}
  {dblp:#1}}
\def\mn@eprint@#1:#2:#3:#4\@nil{\def\@tempa {#1}\def\@tempb {#2}\def\@tempc
  {#3}\ifx \@tempc \@empty \let \@tempc \@tempb \let \@tempb \@tempa \fi \ifx
  \@tempb \@empty \def\@tempb {arXiv}\fi \@ifundefined
  {mn@eprint@\@tempb}{\@tempb:\@tempc}{\expandafter \expandafter \csname
  mn@eprint@\@tempb\endcsname \expandafter{\@tempc}}}

\bibitem[\protect\citeauthoryear{{Binney} \& {Tremaine}}{{Binney} \&
  {Tremaine}}{2008}]{binneytremaine}
{Binney} J.,  {Tremaine} S.,  2008, {Galactic Dynamics: Second Edition}.
Princeton University Press

\bibitem[\protect\citeauthoryear{{Birkin} et~al.,}{{Birkin}
  et~al.}{2024}]{birkin2024}
{Birkin} J.~E.,  et~al., 2024, \mn@doi [\mnras] {10.1093/mnras/stae1089}, \href
  {https://ui.adsabs.harvard.edu/abs/2024MNRAS.531...61B} {531, 61}

\bibitem[\protect\citeauthoryear{{Bouch{\'e}}, {Carfantan}, {Schroetter},
  {Michel-Dansac}  \& {Contini}}{{Bouch{\'e}} et~al.}{2015a}]{galpakcode2015}
{Bouch{\'e}} N.,  {Carfantan} H.,  {Schroetter} I.,  {Michel-Dansac} L.,
  {Contini} T.,  2015a, {GalPaK 3D: Galaxy parameters and kinematics extraction
  from 3D data}, Astrophysics Source Code Library, record ascl:1501.014

\bibitem[\protect\citeauthoryear{{Bouch{\'e}}, {Carfantan}, {Schroetter},
  {Michel-Dansac}  \& {Contini}}{{Bouch{\'e}} et~al.}{2015b}]{bouche2015}
{Bouch{\'e}} N.,  {Carfantan} H.,  {Schroetter} I.,  {Michel-Dansac} L.,
  {Contini} T.,  2015b, \mn@doi [\aj] {10.1088/0004-6256/150/3/92}, \href
  {https://ui.adsabs.harvard.edu/abs/2015AJ....150...92B} {150, 92}

\bibitem[\protect\citeauthoryear{{CASA Team} et~al.,}{{CASA Team}
  et~al.}{2022}]{casa}
{CASA Team} et~al., 2022, \mn@doi [\pasp] {10.1088/1538-3873/ac9642}, \href
  {https://ui.adsabs.harvard.edu/abs/2022PASP..134k4501C} {134, 114501}

\bibitem[\protect\citeauthoryear{{Carilli} \& {Walter}}{{Carilli} \&
  {Walter}}{2013}]{CarilliWalter2013}
{Carilli} C.~L.,  {Walter} F.,  2013, \mn@doi [\araa]
  {10.1146/annurev-astro-082812-140953}, \href
  {https://ui.adsabs.harvard.edu/abs/2013ARA&A..51..105C} {51, 105}

\bibitem[\protect\citeauthoryear{{Conselice}}{{Conselice}}{2014}]{conselice2014}
{Conselice} C.~J.,  2014, \mn@doi [\araa]
  {10.1146/annurev-astro-081913-040037}, \href
  {https://ui.adsabs.harvard.edu/abs/2014ARA&A..52..291C} {52, 291}

\bibitem[\protect\citeauthoryear{{Davies} et~al.,}{{Davies}
  et~al.}{2011}]{davis2011}
{Davies} R.,  et~al., 2011, \mn@doi [\apj] {10.1088/0004-637X/741/2/69}, \href
  {https://ui.adsabs.harvard.edu/abs/2011ApJ...741...69D} {741, 69}

\bibitem[\protect\citeauthoryear{{Davis}, {Bureau}, {Cappellari}, {Sarzi}  \&
  {Blitz}}{{Davis} et~al.}{2013}]{davis2013}
{Davis} T.~A.,  {Bureau} M.,  {Cappellari} M.,  {Sarzi} M.,   {Blitz} L.,
  2013, \mn@doi [\nat] {10.1038/nature11819}, \href
  {https://ui.adsabs.harvard.edu/abs/2013Natur.494..328D} {494, 328}

\bibitem[\protect\citeauthoryear{{Davis}, {Bureau}, {Onishi}, {Cappellari},
  {Iguchi}  \& {Sarzi}}{{Davis} et~al.}{2017}]{davis2017}
{Davis} T.~A.,  {Bureau} M.,  {Onishi} K.,  {Cappellari} M.,  {Iguchi} S.,
  {Sarzi} M.,  2017, \mn@doi [\mnras] {10.1093/mnras/stw3217}, \href
  {https://ui.adsabs.harvard.edu/abs/2017MNRAS.468.4675D} {468, 4675}

\bibitem[\protect\citeauthoryear{{Dekel}, {Sari}  \& {Ceverino}}{{Dekel}
  et~al.}{2009}]{dekel2009}
{Dekel} A.,  {Sari} R.,   {Ceverino} D.,  2009, \mn@doi [\apj]
  {10.1088/0004-637X/703/1/785}, \href
  {https://ui.adsabs.harvard.edu/abs/2009ApJ...703..785D} {703, 785}

\bibitem[\protect\citeauthoryear{{Dessauges-Zavadsky}
  et~al.,}{{Dessauges-Zavadsky} et~al.}{2019}]{dessauges-zavadsky2019}
{Dessauges-Zavadsky} M.,  et~al., 2019, \mn@doi [Nature Astronomy]
  {10.1038/s41550-019-0874-0}, \href
  {https://ui.adsabs.harvard.edu/abs/2019NatAs...3.1115D} {3, 1115}

\bibitem[\protect\citeauthoryear{{Devereaux} et~al.,}{{Devereaux}
  et~al.}{2024}]{devereaux2024}
{Devereaux} T.,  et~al., 2024, \mn@doi [\aap] {10.1051/0004-6361/202348511},
  \href {https://ui.adsabs.harvard.edu/abs/2024A&A...686A.156D} {686, A156}

\bibitem[\protect\citeauthoryear{{Di Teodoro} \& {Fraternali}}{{Di Teodoro} \&
  {Fraternali}}{2015}]{diteodoro2015}
{Di Teodoro} E.~M.,  {Fraternali} F.,  2015, \mn@doi [\mnras]
  {10.1093/mnras/stv1213}, \href
  {https://ui.adsabs.harvard.edu/abs/2015MNRAS.451.3021D} {451, 3021}

\bibitem[\protect\citeauthoryear{{Di Teodoro} \& {Peek}}{{Di Teodoro} \&
  {Peek}}{2021}]{diteodoro2021}
{Di Teodoro} E.~M.,  {Peek} J.~E.~G.,  2021, \mn@doi [\apj]
  {10.3847/1538-4357/ac2cbd}, \href
  {https://ui.adsabs.harvard.edu/abs/2021ApJ...923..220D} {923, 220}

\bibitem[\protect\citeauthoryear{{F{\"o}rster Schreiber} \&
  {Wuyts}}{{F{\"o}rster Schreiber} \& {Wuyts}}{2020}]{forsterschreiber2020}
{F{\"o}rster Schreiber} N.~M.,  {Wuyts} S.,  2020, \mn@doi [\araa]
  {10.1146/annurev-astro-032620-021910}, \href
  {https://ui.adsabs.harvard.edu/abs/2020ARA&A..58..661F} {58, 661}

\bibitem[\protect\citeauthoryear{{F{\"o}rster Schreiber} et~al.,}{{F{\"o}rster
  Schreiber} et~al.}{2009}]{forster2009}
{F{\"o}rster Schreiber} N.~M.,  et~al., 2009, \mn@doi [\apj]
  {10.1088/0004-637X/706/2/1364}, \href
  {https://ui.adsabs.harvard.edu/abs/2009ApJ...706.1364F} {706, 1364}

\bibitem[\protect\citeauthoryear{{F{\"o}rster Schreiber} et~al.,}{{F{\"o}rster
  Schreiber} et~al.}{2018a}]{forster2018}
{F{\"o}rster Schreiber} N.~M.,  et~al., 2018a, \mn@doi [\apjs]
  {10.3847/1538-4365/aadd49}, \href
  {https://ui.adsabs.harvard.edu/abs/2018ApJS..238...21F} {238, 21}

\bibitem[\protect\citeauthoryear{{F{\"o}rster Schreiber} et~al.,}{{F{\"o}rster
  Schreiber} et~al.}{2018b}]{fs2018}
{F{\"o}rster Schreiber} N.~M.,  et~al., 2018b, \mn@doi [\apjs]
  {10.3847/1538-4365/aadd49}, \href
  {https://ui.adsabs.harvard.edu/abs/2018ApJS..238...21F} {238, 21}

\bibitem[\protect\citeauthoryear{{Freeman}}{{Freeman}}{1970}]{freeman1970}
{Freeman} K.~C.,  1970, \mn@doi [\apj] {10.1086/150474}, \href
  {https://ui.adsabs.harvard.edu/abs/1970ApJ...160..811F} {160, 811}

\bibitem[\protect\citeauthoryear{{Genzel} et~al.,}{{Genzel}
  et~al.}{2006}]{genzel2006}
{Genzel} R.,  et~al., 2006, \mn@doi [\nat] {10.1038/nature05052}, \href
  {https://ui.adsabs.harvard.edu/abs/2006Natur.442..786G} {442, 786}

\bibitem[\protect\citeauthoryear{{Genzel} et~al.,}{{Genzel}
  et~al.}{2011}]{genzel2011}
{Genzel} R.,  et~al., 2011, \mn@doi [\apj] {10.1088/0004-637X/733/2/101}, \href
  {https://ui.adsabs.harvard.edu/abs/2011ApJ...733..101G} {733, 101}

\bibitem[\protect\citeauthoryear{{Genzel} et~al.,}{{Genzel}
  et~al.}{2013}]{genzel2013}
{Genzel} R.,  et~al., 2013, \mn@doi [\apj] {10.1088/0004-637X/773/1/68}, \href
  {https://ui.adsabs.harvard.edu/abs/2013ApJ...773...68G} {773, 68}

\bibitem[\protect\citeauthoryear{{Genzel} et~al.,}{{Genzel}
  et~al.}{2023}]{genzel2023}
{Genzel} R.,  et~al., 2023, \mn@doi [\apj] {10.3847/1538-4357/acef1a}, \href
  {https://ui.adsabs.harvard.edu/abs/2023ApJ...957...48G} {957, 48}

\bibitem[\protect\citeauthoryear{{Glazebrook}}{{Glazebrook}}{2013}]{glazebrook2013}
{Glazebrook} K.,  2013, \mn@doi [\pasa] {10.1017/pasa.2013.34}, \href
  {https://ui.adsabs.harvard.edu/abs/2013PASA...30...56G} {30, e056}

\bibitem[\protect\citeauthoryear{{Herrera-Camus} et~al.,}{{Herrera-Camus}
  et~al.}{2022}]{herreracamus2022}
{Herrera-Camus} R.,  et~al., 2022, \mn@doi [\aap]
  {10.1051/0004-6361/202142562}, \href
  {https://ui.adsabs.harvard.edu/abs/2022A&A...665L...8H} {665, L8}

\bibitem[\protect\citeauthoryear{{Johnson} et~al.,}{{Johnson}
  et~al.}{2018}]{johson2018}
{Johnson} H.~L.,  et~al., 2018, \mn@doi [\mnras] {10.1093/mnras/stx3016}, \href
  {https://ui.adsabs.harvard.edu/abs/2018MNRAS.474.5076J} {474, 5076}

\bibitem[\protect\citeauthoryear{{Jones} et~al.,}{{Jones}
  et~al.}{2021}]{jones2021}
{Jones} G.~C.,  et~al., 2021, \mn@doi [\mnras] {10.1093/mnras/stab2226}, \href
  {https://ui.adsabs.harvard.edu/abs/2021MNRAS.507.3540J} {507, 3540}

\bibitem[\protect\citeauthoryear{{Jones} et~al.,}{{Jones}
  et~al.}{2024}]{jones2024}
{Jones} G.~C.,  et~al., 2024, \mn@doi [arXiv e-prints]
  {10.48550/arXiv.2405.12955}, \href
  {https://ui.adsabs.harvard.edu/abs/2024arXiv240512955J} {p. arXiv:2405.12955}

\bibitem[\protect\citeauthoryear{{J{\'o}zsa}, {Kenn}, {Klein}  \&
  {Oosterloo}}{{J{\'o}zsa} et~al.}{2007}]{jozsa2007}
{J{\'o}zsa} G.~I.~G.,  {Kenn} F.,  {Klein} U.,   {Oosterloo} T.~A.,  2007,
  \mn@doi [\aap] {10.1051/0004-6361:20066164}, \href
  {https://ui.adsabs.harvard.edu/abs/2007A&A...468..731J} {468, 731}

\bibitem[\protect\citeauthoryear{{Kade}, {Knudsen}, {Bewketu Belete}, {Yang},
  {K{\"o}nig}, {Stanley}  \& {Scholtz}}{{Kade} et~al.}{2024}]{kade2024}
{Kade} K.,  {Knudsen} K.~K.,  {Bewketu Belete} A.,  {Yang} C.,  {K{\"o}nig} S.,
   {Stanley} F.,   {Scholtz} J.,  2024, \mn@doi [\aap]
  {10.1051/0004-6361/202347453}, \href
  {https://ui.adsabs.harvard.edu/abs/2024A&A...684A..56K} {684, A56}

\bibitem[\protect\citeauthoryear{{Kohandel}, {Pallottini}, {Ferrara},
  {Zanella}, {Behrens}, {Carniani}, {Gallerani}  \& {Vallini}}{{Kohandel}
  et~al.}{2019}]{kohandel2019}
{Kohandel} M.,  {Pallottini} A.,  {Ferrara} A.,  {Zanella} A.,  {Behrens} C.,
  {Carniani} S.,  {Gallerani} S.,   {Vallini} L.,  2019, \mn@doi [\mnras]
  {10.1093/mnras/stz1486}, \href
  {https://ui.adsabs.harvard.edu/abs/2019MNRAS.487.3007K} {487, 3007}

\bibitem[\protect\citeauthoryear{{Kohandel}, {Pallottini}, {Ferrara},
  {Carniani}, {Gallerani}, {Vallini}, {Zanella}  \& {Behrens}}{{Kohandel}
  et~al.}{2020}]{kohandel2020}
{Kohandel} M.,  {Pallottini} A.,  {Ferrara} A.,  {Carniani} S.,  {Gallerani}
  S.,  {Vallini} L.,  {Zanella} A.,   {Behrens} C.,  2020, \mn@doi [\mnras]
  {10.1093/mnras/staa2792}, \href
  {https://ui.adsabs.harvard.edu/abs/2020MNRAS.499.1250K} {499, 1250}

\bibitem[\protect\citeauthoryear{{Kohandel}, {Pallottini}, {Ferrara},
  {Zanella}, {Rizzo}  \& {Carniani}}{{Kohandel} et~al.}{2024}]{kohandel2024}
{Kohandel} M.,  {Pallottini} A.,  {Ferrara} A.,  {Zanella} A.,  {Rizzo} F.,
  {Carniani} S.,  2024, \mn@doi [\aap] {10.1051/0004-6361/202348209}, \href
  {https://ui.adsabs.harvard.edu/abs/2024A&A...685A..72K} {685, A72}

\bibitem[\protect\citeauthoryear{{Krumholz} \& {Burkert}}{{Krumholz} \&
  {Burkert}}{2010}]{krumholtz2010}
{Krumholz} M.,  {Burkert} A.,  2010, \mn@doi [\apj]
  {10.1088/0004-637X/724/2/895}, \href
  {https://ui.adsabs.harvard.edu/abs/2010ApJ...724..895K} {724, 895}

\bibitem[\protect\citeauthoryear{{Lamperti} et~al.,}{{Lamperti}
  et~al.}{2024}]{lamperti2024}
{Lamperti} I.,  et~al., 2024, \mn@doi [\aap] {10.1051/0004-6361/202451021},
  \href {https://ui.adsabs.harvard.edu/abs/2024A&A...691A.153L} {691, A153}

\bibitem[\protect\citeauthoryear{{Law}, {Steidel}, {Erb}, {Larkin}, {Pettini},
  {Shapley}  \& {Wright}}{{Law} et~al.}{2009}]{law2009}
{Law} D.~R.,  {Steidel} C.~C.,  {Erb} D.~K.,  {Larkin} J.~E.,  {Pettini} M.,
  {Shapley} A.~E.,   {Wright} S.~A.,  2009, \mn@doi [\apj]
  {10.1088/0004-637X/697/2/2057}, \href
  {https://ui.adsabs.harvard.edu/abs/2009ApJ...697.2057L} {697, 2057}

\bibitem[\protect\citeauthoryear{{Lee} et~al.,}{{Lee} et~al.}{2025}]{lee2024}
{Lee} L.~L.,  et~al., 2025, \mn@doi [\apj] {10.3847/1538-4357/ad90b5}, \href
  {https://ui.adsabs.harvard.edu/abs/2025ApJ...978...14L} {978, 14}

\bibitem[\protect\citeauthoryear{{Levy} et~al.,}{{Levy}
  et~al.}{2018}]{levy2018}
{Levy} R.~C.,  et~al., 2018, \mn@doi [\apj] {10.3847/1538-4357/aac2e5}, \href
  {https://ui.adsabs.harvard.edu/abs/2018ApJ...860...92L} {860, 92}

\bibitem[\protect\citeauthoryear{{Molina}, {Ibar}, {Smail}, {Swinbank},
  {Villard}, {Escala}, {Sobral}  \& {Hughes}}{{Molina}
  et~al.}{2019}]{molina2019}
{Molina} J.,  {Ibar} E.,  {Smail} I.,  {Swinbank} A.~M.,  {Villard} E.,
  {Escala} A.,  {Sobral} D.,   {Hughes} T.~M.,  2019, \mn@doi [\mnras]
  {10.1093/mnras/stz1643}, \href
  {https://ui.adsabs.harvard.edu/abs/2019MNRAS.487.4856M} {487, 4856}

\bibitem[\protect\citeauthoryear{{Motta} et~al.,}{{Motta}
  et~al.}{2018}]{motta2018}
{Motta} V.,  et~al., 2018, \mn@doi [\apjl] {10.3847/2041-8213/aad6de}, \href
  {https://ui.adsabs.harvard.edu/abs/2018ApJ...863L..16M} {863, L16}

\bibitem[\protect\citeauthoryear{{Neeleman}, {Prochaska}, {Kanekar}  \&
  {Rafelski}}{{Neeleman} et~al.}{2020}]{neeleman2020thecode}
{Neeleman} M.,  {Prochaska} J.~X.,  {Kanekar} N.,   {Rafelski} M.,  2020,
  {qubefit: MCMC kinematic modeling}, Astrophysics Source Code Library, record
  ascl:2005.013 (\mn@eprint {ascl} {2005.013})

\bibitem[\protect\citeauthoryear{{Neeleman} et~al.,}{{Neeleman}
  et~al.}{2021}]{neeleman2021}
{Neeleman} M.,  et~al., 2021, \mn@doi [\apj] {10.3847/1538-4357/abe70f}, \href
  {https://ui.adsabs.harvard.edu/abs/2021ApJ...911..141N} {911, 141}

\bibitem[\protect\citeauthoryear{{Pillepich} et~al.,}{{Pillepich}
  et~al.}{2019}]{pillepich2019}
{Pillepich} A.,  et~al., 2019, \mn@doi [\mnras] {10.1093/mnras/stz2338}, \href
  {https://ui.adsabs.harvard.edu/abs/2019MNRAS.490.3196P} {490, 3196}

\bibitem[\protect\citeauthoryear{{Planck Collaboration} et~al.,}{{Planck
  Collaboration} et~al.}{2016}]{planck15cosmology}
{Planck Collaboration} et~al., 2016, \mn@doi [\aap]
  {10.1051/0004-6361/201525830}, \href
  {https://ui.adsabs.harvard.edu/abs/2016A&A...594A..13P} {594, A13}

\bibitem[\protect\citeauthoryear{{Pope} et~al.,}{{Pope}
  et~al.}{2023}]{pope2023}
{Pope} A.,  et~al., 2023, \mn@doi [\apjl] {10.3847/2041-8213/acdf5a}, \href
  {https://ui.adsabs.harvard.edu/abs/2023ApJ...951L..46P} {951, L46}

\bibitem[\protect\citeauthoryear{{Price} et~al.,}{{Price}
  et~al.}{2021}]{price2021}
{Price} S.~H.,  et~al., 2021, \mn@doi [\apj] {10.3847/1538-4357/ac22ad}, \href
  {https://ui.adsabs.harvard.edu/abs/2021ApJ...922..143P} {922, 143}

\bibitem[\protect\citeauthoryear{{Rizzo}, {Vegetti}, {Powell}, {Fraternali},
  {McKean}, {Stacey}  \& {White}}{{Rizzo} et~al.}{2020}]{rizzo2020}
{Rizzo} F.,  {Vegetti} S.,  {Powell} D.,  {Fraternali} F.,  {McKean} J.~P.,
  {Stacey} H.~R.,   {White} S.~D.~M.,  2020, \mn@doi [\nat]
  {10.1038/s41586-020-2572-6}, \href
  {https://ui.adsabs.harvard.edu/abs/2020Natur.584..201R} {584, 201}

\bibitem[\protect\citeauthoryear{{Rizzo}, {Vegetti}, {Fraternali}, {Stacey}  \&
  {Powell}}{{Rizzo} et~al.}{2021}]{rizzo2021}
{Rizzo} F.,  {Vegetti} S.,  {Fraternali} F.,  {Stacey} H.~R.,   {Powell} D.,
  2021, \mn@doi [\mnras] {10.1093/mnras/stab2295}, \href
  {https://ui.adsabs.harvard.edu/abs/2021MNRAS.507.3952R} {507, 3952}

\bibitem[\protect\citeauthoryear{{Rizzo}, {Kohandel}, {Pallottini}, {Zanella},
  {Ferrara}, {Vallini}  \& {Toft}}{{Rizzo} et~al.}{2022}]{rizzo2022}
{Rizzo} F.,  {Kohandel} M.,  {Pallottini} A.,  {Zanella} A.,  {Ferrara} A.,
  {Vallini} L.,   {Toft} S.,  2022, \mn@doi [\aap]
  {10.1051/0004-6361/202243582}, \href
  {https://ui.adsabs.harvard.edu/abs/2022A&A...667A...5R} {667, A5}

\bibitem[\protect\citeauthoryear{{Rizzo} et~al.,}{{Rizzo}
  et~al.}{2023}]{rizzo2023}
{Rizzo} F.,  et~al., 2023, \mn@doi [\aap] {10.1051/0004-6361/202346444}, \href
  {https://ui.adsabs.harvard.edu/abs/2023A&A...679A.129R} {679, A129}

\bibitem[\protect\citeauthoryear{{Robertson} et~al.,}{{Robertson}
  et~al.}{2023}]{robertson2023}
{Robertson} B.~E.,  et~al., 2023, \mn@doi [\apjl] {10.3847/2041-8213/aca086},
  \href {https://ui.adsabs.harvard.edu/abs/2023ApJ...942L..42R} {942, L42}

\bibitem[\protect\citeauthoryear{{Rogstad}, {Lockhart}  \& {Wright}}{{Rogstad}
  et~al.}{1974}]{rogstad1974}
{Rogstad} D.~H.,  {Lockhart} I.~A.,   {Wright} M.~C.~H.,  1974, \mn@doi [\apj]
  {10.1086/153164}, \href
  {https://ui.adsabs.harvard.edu/abs/1974ApJ...193..309R} {193, 309}

\bibitem[\protect\citeauthoryear{{Roman-Oliveira}, {Fraternali}  \&
  {Rizzo}}{{Roman-Oliveira} et~al.}{2023}]{romanoliveira2023}
{Roman-Oliveira} F.,  {Fraternali} F.,   {Rizzo} F.,  2023, \mn@doi [\mnras]
  {10.1093/mnras/stad530}, \href
  {https://ui.adsabs.harvard.edu/abs/2023MNRAS.521.1045R} {521, 1045}

\bibitem[\protect\citeauthoryear{{Rowland} et~al.,}{{Rowland}
  et~al.}{2024}]{rowland2024}
{Rowland} L.~E.,  et~al., 2024, \mn@doi [\mnras] {10.1093/mnras/stae2217},
  \href {https://ui.adsabs.harvard.edu/abs/2024MNRAS.535.2068R} {535, 2068}

\bibitem[\protect\citeauthoryear{{Rubin} \& {Ford}}{{Rubin} \&
  {Ford}}{1970}]{rubin1970}
{Rubin} V.~C.,  {Ford} W.~Kent J.,  1970, \mn@doi [\apj] {10.1086/150317},
  \href {https://ui.adsabs.harvard.edu/abs/1970ApJ...159..379R} {159, 379}

\bibitem[\protect\citeauthoryear{{Scholtz} et~al.,}{{Scholtz}
  et~al.}{2025a}]{scholtz2025}
{Scholtz} J.,  et~al., 2025a, \mn@doi [arXiv e-prints]
  {10.48550/arXiv.2503.10751}, \href
  {https://ui.adsabs.harvard.edu/abs/2025arXiv250310751S} {p. arXiv:2503.10751}

\bibitem[\protect\citeauthoryear{{Scholtz} et~al.,}{{Scholtz}
  et~al.}{2025b}]{scholtz2024}
{Scholtz} J.,  et~al., 2025b, \mn@doi [\mnras] {10.1093/mnras/staf518}, \href
  {https://ui.adsabs.harvard.edu/abs/2025MNRAS.539.2463S} {539, 2463}

\bibitem[\protect\citeauthoryear{{S{\'e}rsic}}{{S{\'e}rsic}}{1963}]{sersic1963}
{S{\'e}rsic} J.~L.,  1963, Boletin de la Asociacion Argentina de Astronomia La
  Plata Argentina, 6, 41

\bibitem[\protect\citeauthoryear{{Simons} et~al.,}{{Simons}
  et~al.}{2019}]{simons2019}
{Simons} R.~C.,  et~al., 2019, \mn@doi [\apj] {10.3847/1538-4357/ab07c9}, \href
  {https://ui.adsabs.harvard.edu/abs/2019ApJ...874...59S} {874, 59}

\bibitem[\protect\citeauthoryear{{Smit} et~al.,}{{Smit}
  et~al.}{2018}]{smit2018}
{Smit} R.,  et~al., 2018, \mn@doi [\nat] {10.1038/nature24631}, \href
  {https://ui.adsabs.harvard.edu/abs/2018Natur.553..178S} {553, 178}

\bibitem[\protect\citeauthoryear{{Solomon}, {Downes}  \& {Radford}}{{Solomon}
  et~al.}{1992}]{solomon1992}
{Solomon} P.~M.,  {Downes} D.,   {Radford} S.~J.~E.,  1992, \mn@doi [\apjl]
  {10.1086/186569}, \href
  {https://ui.adsabs.harvard.edu/abs/1992ApJ...398L..29S} {398, L29}

\bibitem[\protect\citeauthoryear{{Stott} et~al.,}{{Stott}
  et~al.}{2016}]{stott2016}
{Stott} J.~P.,  et~al., 2016, \mn@doi [\mnras] {10.1093/mnras/stw129}, \href
  {https://ui.adsabs.harvard.edu/abs/2016MNRAS.457.1888S} {457, 1888}

\bibitem[\protect\citeauthoryear{{Swinbank}, {Sobral}, {Smail}, {Geach},
  {Best}, {McCarthy}, {Crain}  \& {Theuns}}{{Swinbank}
  et~al.}{2012}]{swinbank2012}
{Swinbank} A.~M.,  {Sobral} D.,  {Smail} I.,  {Geach} J.~E.,  {Best} P.~N.,
  {McCarthy} I.~G.,  {Crain} R.~A.,   {Theuns} T.,  2012, \mn@doi [\mnras]
  {10.1111/j.1365-2966.2012.21774.x}, \href
  {https://ui.adsabs.harvard.edu/abs/2012MNRAS.426..935S} {426, 935}

\bibitem[\protect\citeauthoryear{{Telikova} et~al.,}{{Telikova}
  et~al.}{2025}]{telikova2024}
{Telikova} K.,  et~al., 2025, \mn@doi [\aap] {10.1051/0004-6361/202452990},
  \href {https://ui.adsabs.harvard.edu/abs/2025A&A...699A...5T} {699, A5}

\bibitem[\protect\citeauthoryear{{Turner} et~al.,}{{Turner}
  et~al.}{2017}]{turner2017}
{Turner} O.~J.,  et~al., 2017, \mn@doi [\mnras] {10.1093/mnras/stx1366}, \href
  {https://ui.adsabs.harvard.edu/abs/2017MNRAS.471.1280T} {471, 1280}

\bibitem[\protect\citeauthoryear{{{\"U}bler} et~al.,}{{{\"U}bler}
  et~al.}{2018}]{ubler2018}
{{\"U}bler} H.,  et~al., 2018, \mn@doi [\apjl] {10.3847/2041-8213/aaacfa},
  \href {https://ui.adsabs.harvard.edu/abs/2018ApJ...854L..24U} {854, L24}

\bibitem[\protect\citeauthoryear{{Wisnioski} et~al.,}{{Wisnioski}
  et~al.}{2015}]{wisnioski2015}
{Wisnioski} E.,  et~al., 2015, \mn@doi [\apj] {10.1088/0004-637X/799/2/209},
  \href {https://ui.adsabs.harvard.edu/abs/2015ApJ...799..209W} {799, 209}

\bibitem[\protect\citeauthoryear{{Wisnioski} et~al.,}{{Wisnioski}
  et~al.}{2019}]{wisnioski2019}
{Wisnioski} E.,  et~al., 2019, \mn@doi [\apj] {10.3847/1538-4357/ab4db8}, \href
  {https://ui.adsabs.harvard.edu/abs/2019ApJ...886..124W} {886, 124}

\bibitem[\protect\citeauthoryear{{Wuyts} et~al.,}{{Wuyts}
  et~al.}{2011}]{wuyts2011b}
{Wuyts} S.,  et~al., 2011, \mn@doi [\apj] {10.1088/0004-637X/742/2/96}, \href
  {https://ui.adsabs.harvard.edu/abs/2011ApJ...742...96W} {742, 96}

\bibitem[\protect\citeauthoryear{{van der Wel} et~al.,}{{van der Wel}
  et~al.}{2014}]{vanderwel2014}
{van der Wel} A.,  et~al., 2014, \mn@doi [\apj] {10.1088/0004-637X/788/1/28},
  \href {https://ui.adsabs.harvard.edu/abs/2014ApJ...788...28V} {788, 28}

\makeatother
\end{thebibliography}



\appendix

\section{Convergence of 3D fits} \label{app:tables:3Dconvergence}
The following three tables lists each of the intrinsic cube setups, as detailed in Table \ref{table:simgals_for_simalma}, together with the setups used for the 3D-kinematical fitting carried out with \barolo, \galpak, and \qubefit. 
These tables show which fits converged according to the tool specific convergence requirements. 
H, M, and L stand for high data quality, medium data quality, and low data quality respectively as synthetically observed using \texttt{simalma}. 
The freeINCL, freePA etc, state if the inclination or position angle is left to vary in the fitting procedure, and if the inclination is fixed to $\pm10$ of its actual value. The flux, rotation velocity, and velocity dispersion are always free parameters in the fits.

\begin{table*}
\caption{\barolo\, 3D-fitting setup and convergence. For rotation dominated systems \barolo's convergence rate is 100\%, for non-ideal disks ('inBetween' and 'dispDom'; $V_{\rm rot}/\sigma_{\rm V}\leq1$) it is 81\%. On the entire sample the convergence rate is 93\%. }
    \makebox[\textwidth][l]{%
    \begin{tabular}{l|p{0.6in}p{0.6in}p{0.6in}p{0.6in}p{0.6in}p{0.6in}|p{0.6in}}
     &    &      &         &      &  &      \\ 
    \cmidrule{2-8}
   Setup & \CenterCell{fixedPA\\\&INCL} & \CenterCell{freeINCL} & \CenterCell{freePA} & \CenterCell{freeINCL\\\&PA} & \CenterCell{incl+10\\freePA} & \CenterCell{incl-10\\freePA} & \CenterCell{converged\\ (H,M,L\%))}\\
    \midrule
    
        exp-risFlat-const & \textcolor{green}{H, M, L} & \textcolor{green}{H, M, L} & \textcolor{green}{H, M, L} & \textcolor{green}{H, M, L} & \textcolor{green}{H, M, L} & \textcolor{green}{H, M, L} & \textcolor{green}{100, 100, 100}\\ 
        
        exp-risFlat-exp & \textcolor{green}{H, M, L} & \textcolor{green}{H, M, L} & \textcolor{green}{H, M, L} & \textcolor{green}{H, M, L} & \textcolor{green}{H, M, L} & \textcolor{green}{H, M, L} & \textcolor{green}{100, 100, 100}\\ 
        
        exp-risDec-const  & \textcolor{green}{H, M, L} & \textcolor{green}{H, M, L} & \textcolor{green}{H, M, L} & \textcolor{green}{H, M, L} & \textcolor{green}{H, M, L} & \textcolor{green}{H, M, L} & \textcolor{green}{100, 100, 100}\\ 

        corelog-risFlat-const & \textcolor{green}{H, M, L} & \textcolor{green}{H, M, L} & \textcolor{green}{H, M, L} & \textcolor{green}{H, M, L} & \textcolor{green}{H, M, L} & \textcolor{green}{H, M, L}  & \textcolor{green}{100, 100, 100}\\ 

        corelog-risFlat-exp & \textcolor{green}{H, M, L} & \textcolor{green}{H, M, L} & \textcolor{green}{H, M, L} & \textcolor{green}{H, M, L} & \textcolor{green}{H, M, L} & \textcolor{green}{H, M, L} & \textcolor{green}{100, 100, 100}\\ 

        rotDom-constDisp & \textcolor{green}{H, M, L} & \textcolor{green}{H, M, L} & \textcolor{green}{H, M, L} & \textcolor{green}{H, M, L} & \textcolor{green}{H, M, L} & \textcolor{green}{H, M, L} & \textcolor{green}{100, 100, 100}\\ 

        rotDom-expDisp & \textcolor{green}{H, M, L} & \textcolor{green}{H, M, L} & \textcolor{green}{H, M, L} & \textcolor{green}{H, M, L} & \textcolor{green}{H, M, L} & \textcolor{green}{H, M, L} & \textcolor{green}{100, 100, 100}\\ 

        inBetween-constDisp & \textcolor{green}{H, M, L} & \textcolor{green}{H, M, L} & \textcolor{green}{H, M, L} & \textcolor{green}{H, M, L} & \textcolor{green}{H, M, L} & \textcolor{green}{H, M, L}  & \textcolor{green}{100, 100, 100}\\ 

        inBetween-expDisp & \textcolor{green}{H, M, L} & \textcolor{green}{H, M, L} & \textcolor{green}{H, M, L} & \textcolor{green}{H, M, L} & \textcolor{green}{H, M, L} & \textcolor{green}{H, M, L}  & \textcolor{green}{100, 100, 100}\\ 

        dispDom-constDisp & \textcolor{green}{H, M, }\textcolor{red}{L} & \textcolor{green}{H, M, }\textcolor{red}{L} & \textcolor{green}{H, M, }\textcolor{red}{L} & \textcolor{green}{H, }\textcolor{red}{M, L} & \textcolor{green}{H, M, }\textcolor{red}{L} & \textcolor{green}{H, M, }\textcolor{red}{L} & \textcolor{green}{100, }\textcolor{orange}{83, }\textcolor{red}{0} \\ 

        dispDom-expDisp & \textcolor{green}{H, M, }\textcolor{red}{L} & \textcolor{green}{H, M, }\textcolor{red}{L} & \textcolor{green}{H, M, }\textcolor{red}{L} & \textcolor{green}{H, }\textcolor{red}{M, L} & \textcolor{green}{H, M, }\textcolor{red}{L} & \textcolor{green}{H, M, }\textcolor{red}{L} & \textcolor{green}{100, }\textcolor{orange}{83, }\textcolor{red}{0} \\ 
        
    \end{tabular}
    }
    
    \label{table:3Dconvergence_barolo}
\end{table*}

\begin{table*}
    \caption{\galpak\, 3D-fitting setup and convergence. For rotation dominated systems \galpak's convergence rate is 92\%, for non-ideal disks ('inBetween' and 'dispDom'; $V_{\rm rot}/\sigma_{\rm V}\leq1$) it is 73\%. On the entire sample the convergence rate is 85\%.}
    \makebox[\textwidth][l]{%
    \begin{tabular}{l|*{6}{p{0.6in}}|p{0.6in}p{0.6in}|p{0.6in}}
    &    &      &    &     &      &  &    &  \\ 
    & \multicolumn{6}{c|}{\CenterCell{exp-arctan-thin}} & \multicolumn{2}{l|}{\CenterCell{exp-arctan-thick}} \\
    \cmidrule{2-9}
    Setup & \CenterCell{fixedPA\\\&INCL} & \CenterCell{freeINCL} & \CenterCell{freePA} & \CenterCell{freeINCL\\\&PA} & \CenterCell{incl+10\\freePA} & \CenterCell{incl-10\\freePA} & \CenterCell{thick\\freeINCL} & \CenterCell{thick\\freePA} & \CenterCell{converged\\ (H,M,L\%))}\\
    \midrule
                        
        exp-risFlat-const & \textcolor{green}{H, M, L} & \textcolor{green}{H, M, L} & \textcolor{green}{H, M, L} & \textcolor{green}{H, M, L} & \textcolor{green}{H, M, L} & \textcolor{green}{H, M, L}  & \textcolor{green}{H, M, L} & \textcolor{green}{H, M, L} & \textcolor{green}{100, 100, 100}\\
        
        exp-risFlat-exp & \textcolor{green}{H, M, L} & \textcolor{red}{H, }\textcolor{green}{M, L} & \textcolor{green}{H, M, L} & \textcolor{red}{H, }\textcolor{green}{M, L} & \textcolor{red}{H, }\textcolor{green}{M, L} & \textcolor{red}{H, }\textcolor{green}{M, L}  & \textcolor{red}{H, }\textcolor{green}{M, L} & \textcolor{red}{H, }\textcolor{green}{M, L} & \textcolor{orange}{25, }\textcolor{green}{100, 100}\\
        
        exp-risDec-const  & \textcolor{green}{H, M, L} & \textcolor{green}{H, M, L} & \textcolor{green}{H, M, L} & \textcolor{green}{H, M, L} & \textcolor{green}{H, M, L} & \textcolor{green}{H, M, L}  & \textcolor{green}{H, M, L} & \textcolor{green}{H, M, L}  & \textcolor{green}{100, 100, 100}\\
        
        corelog-risFlat-const & \textcolor{green}{H, M, L} & \textcolor{green}{H, M, L} & \textcolor{green}{H, M, L} & \textcolor{green}{H, M, L} & \textcolor{green}{H, M, L} & \textcolor{green}{H, M, L}  & \textcolor{green}{H, M, L} & \textcolor{green}{H, M, L}  & \textcolor{green}{100, 100, 100}\\

        corelog-risFlat-exp & \textcolor{red}{H, }\textcolor{green}{M, L}  & \textcolor{red}{H, }\textcolor{green}{M, L} & \textcolor{green}{H, M, L} & \textcolor{red}{H, }\textcolor{green}{M, L} & \textcolor{red}{H, }\textcolor{green}{M, L} & \textcolor{red}{H, }\textcolor{green}{M, L}  & \textcolor{red}{H, }\textcolor{green}{M, L} & \textcolor{red}{H, }\textcolor{green}{M, L}  & \textcolor{orange}{13, }\textcolor{green}{100, 100}\\

        rotDom-constDisp & \textcolor{green}{H, M, } & \textcolor{green}{H, M, L} & \textcolor{green}{H, M, L} & \textcolor{green}{H, M, L} & \textcolor{green}{H, M, L}  & \textcolor{green}{H, M, L} & \textcolor{green}{H, M, L} & \textcolor{green}{H, M, L}  & \textcolor{green}{100, 100, 100}\\

        rotDom-expDisp & \textcolor{green}{H, M, L} & \textcolor{green}{H, M, L} & \textcolor{green}{H, M, L} & \textcolor{green}{H, M, L} & \textcolor{green}{H, M, L} & \textcolor{green}{H, M, L} & \textcolor{green}{H, M, L} & \textcolor{green}{H, M, L}  & \textcolor{green}{100, 100, 100}\\

        inBetween-constDisp & \textcolor{green}{H, M, L} & \textcolor{green}{H, M, L} & \textcolor{green}{H, M, L} & \textcolor{green}{H, M, L} & \textcolor{green}{H, M, L} & \textcolor{green}{H, M, L}  & \textcolor{green}{H, M, L} & \textcolor{green}{H, M, L}  & \textcolor{green}{100, 100, 100}\\

        inBetween-expDisp & \textcolor{red}{H, }\textcolor{green}{M, L}  & \textcolor{red}{H, }\textcolor{green}{M, L} & \textcolor{red}{H, M, }\textcolor{green}{L} & \textcolor{red}{H, }\textcolor{green}{M, L} & \textcolor{red}{H, }\textcolor{green}{M, L} & \textcolor{red}{H, }\textcolor{green}{M, L}  & \textcolor{red}{H, }\textcolor{green}{M, L} & \textcolor{red}{H, }\textcolor{green}{M, L}  & \textcolor{red}{0, }\textcolor{orange}{88, }\textcolor{green}{100}\\

        dispDom-constDisp & \textcolor{red}{H, }\textcolor{green}{M, L} & \textcolor{red}{H, }\textcolor{green}{M, L} & \textcolor{red}{H, }\textcolor{green}{M, L} & \textcolor{red}{H, }\textcolor{green}{M, L} & \textcolor{red}{H, }\textcolor{green}{M, L} & \textcolor{red}{H, }\textcolor{green}{M, L} & \textcolor{red}{H, }\textcolor{green}{M, L} & \textcolor{red}{H, }\textcolor{green}{M, L}  & \textcolor{red}{0, }\textcolor{green}{100, 100}\\

        dispDom-expDisp & \textcolor{red}{H, }\textcolor{green}{M, L}  & \textcolor{red}{H, }\textcolor{green}{M, L} & \textcolor{red}{H, }\textcolor{green}{M, L} & \textcolor{red}{H, }\textcolor{green}{M, L} & \textcolor{red}{H, }\textcolor{green}{M, L} & \textcolor{red}{H, }\textcolor{green}{M, L}  & \textcolor{red}{H, M, }\textcolor{green}{L} & \textcolor{red}{H, }\textcolor{green}{M, L}  & \textcolor{red}{0, }\textcolor{orange}{88, }\textcolor{green}{100}\\
        
    \end{tabular}
    \label{table:3Dconvergence_galpak}
    }
\end{table*}

\begin{table*}
    \caption{\qubefit\, 3D-fitting setup and convergence. For rotation dominated systems \qubefit's convergence rate is 99\%, for non-ideal disks ('inBetween' and 'dispDom'; $V_{\rm rot}/\sigma_{\rm V}\leq1$) it is 46\%. On the entire sample the convergence rate is 80\%.}
\label{table:3Dconvergence_qubefit}
\makebox[\textwidth][l]{%
\begin{tabular}{l|*{6}{p{0.6in}}|p{0.6in}|p{0.6in}|p{0.6in}}
    &    &      &    &     &      &  &    &  \\ 
    & \multicolumn{6}{c|}{\CenterCell{ }} & \CenterCell{\hspace{0.1cm} exp-} & \CenterCell{\hspace{0.1cm} exp-} \\
    & \multicolumn{6}{c|}{\CenterCell{exp-arctan-const}} & \CenterCell{-const-} & \CenterCell{-arctan-} \\
    & \multicolumn{6}{c|}{\CenterCell{ }} & \CenterCell{\hspace{0.05cm} -const} & \CenterCell{\hspace{0.1cm} -exp} \\
    \cmidrule{2-9}
    Setup & \CenterCell{fixedPA\\\&INCL} & \CenterCell{freeINCL} & \CenterCell{freePA} & \CenterCell{freeINCL\\\&PA} & \CenterCell{incl+10\\freePA} & \CenterCell{incl-10\\freePA} & \CenterCell{freePA} & \CenterCell{freePA} & \CenterCell{converged\\ (H,M,L\%))}\\
    \midrule
                        
        exp-risFlat-const & \textcolor{green}{H, M, L} & \textcolor{green}{H, M, L} & \textcolor{green}{H, M, L} & \textcolor{green}{H, M, L} & \textcolor{green}{H, M, L} & \textcolor{green}{H, M, L} & \textcolor{green}{H, M, L} & \textcolor{green}{H, M, L} & \textcolor{green}{100, 100, 100}\\
        
        exp-risFlat-exp & \textcolor{green}{H, M, L} & \textcolor{green}{H, M, L} & \textcolor{green}{H, M, L} & \textcolor{green}{H, M, L} & \textcolor{green}{H, M, L} & \textcolor{green}{H, M, L} & \textcolor{green}{H, M, L} & \textcolor{green}{H, M, L} & \textcolor{green}{100, 100, 100}\\
        
        exp-risDec-const  & \textcolor{green}{H, M, L} & \textcolor{green}{H, M, L} & \textcolor{green}{H, M, L} & \textcolor{green}{H, M, L} & \textcolor{green}{H, M, L} & \textcolor{green}{H, M, L} & \textcolor{green}{H, M, L} & \textcolor{green}{H, M, L} & \textcolor{green}{100, 100, 100}\\
        
        corelog-risFlat-const & \textcolor{green}{H, M, L} & \textcolor{green}{H, M, L} & \textcolor{green}{H, M, L} & \textcolor{green}{H, M, L} & \textcolor{green}{H, M, L} & \textcolor{green}{H, M, L} & \textcolor{green}{H, M, L} & \textcolor{green}{H, M, L} & \textcolor{green}{100, 100, 100}\\

        corelog-risFlat-exp & \textcolor{green}{H, M, L} & \textcolor{green}{H, M, L} & \textcolor{green}{H, M, L} & \textcolor{green}{H, M, L} & \textcolor{green}{H, M, }\textcolor{red}{L} & \textcolor{green}{H, M, L} & \textcolor{green}{H, M, L} & \textcolor{green}{H, M, L} & \textcolor{green}{100, 100, }\textcolor{orange}{88}\\

        rotDom-constDisp & \textcolor{green}{H, M, L} & \textcolor{green}{H, M, L} & \textcolor{green}{H, M, L} & \textcolor{green}{H, M, L} & \textcolor{green}{H, M, L} & \textcolor{green}{H, M, L} & \textcolor{green}{H, M, L} & \textcolor{green}{H, M, L} & \textcolor{green}{100, 100, 100}\\

        rotDom-expDisp & \textcolor{green}{H, M, L} & \textcolor{green}{H, M, L} & \textcolor{green}{H, M, L} & \textcolor{green}{H, M, L} & \textcolor{green}{H, M, L} & \textcolor{green}{H, M, L} & \textcolor{green}{H, M, L} & \textcolor{green}{H, M, L} & \textcolor{green}{100, 100, 100}\\

        inBetween-constDisp & \textcolor{green}{H, M, }\textcolor{red}{L} & \textcolor{green}{H, M, }\textcolor{red}{L} & \textcolor{green}{H, M, }\textcolor{red}{L} & \textcolor{green}{H, M, }\textcolor{red}{L} & \textcolor{green}{H, M, }\textcolor{red}{L} & \textcolor{green}{H, M, }\textcolor{red}{L} & \textcolor{green}{H, M, L} & \textcolor{green}{H, M, }\textcolor{red}{L} & \textcolor{green}{100, 100, }\textcolor{orange}{13}\\

        inBetween-expDisp & \textcolor{green}{H, M, }\textcolor{red}{L} & \textcolor{green}{H, M, }\textcolor{red}{L} & \textcolor{green}{H, M, }\textcolor{red}{L} & \textcolor{green}{H, M, }\textcolor{red}{L} & \textcolor{green}{H, M, }\textcolor{red}{L} & \textcolor{red}{H, }\textcolor{green}{M, }\textcolor{red}{L} & \textcolor{green}{H, M, L} & \textcolor{red}{H, }\textcolor{green}{M, L}  & \textcolor{orange}{75, }\textcolor{green}{100, }\textcolor{orange}{25}\\

        dispDom-constDisp & \textcolor{red}{H, M, L} & \textcolor{red}{H, M, L} & \textcolor{red}{H, M, L} & \textcolor{red}{H, M, L} & \textcolor{red}{H, }\textcolor{green}{M, }\textcolor{red}{L} & \textcolor{red}{H, M, L} & \textcolor{green}{H, M, }\textcolor{red}{L} & \textcolor{green}{H, }\textcolor{red}{M, L} & \textcolor{orange}{25, 25, }\textcolor{red}{0}\\

        dispDom-expDisp & \textcolor{red}{H, M, L} & \textcolor{red}{H, M, L} & \textcolor{red}{H, }\textcolor{green}{M, }\textcolor{red}{L} & \textcolor{red}{H, M, L} & \textcolor{red}{H, M, L} & \textcolor{red}{H, }\textcolor{green}{M, }\textcolor{red}{L} & \textcolor{green}{H, M, L} & \textcolor{green}{H, M, }\textcolor{red}{L} & \textcolor{orange}{25, 50, 13}\\
        
    \end{tabular}
    }
\end{table*}


\section{The effect of SNR cutoff with \barolo} \label{app:barolo_SNR_section}

\begin{figure*}
    \centering
    \includegraphics[width=\textwidth]{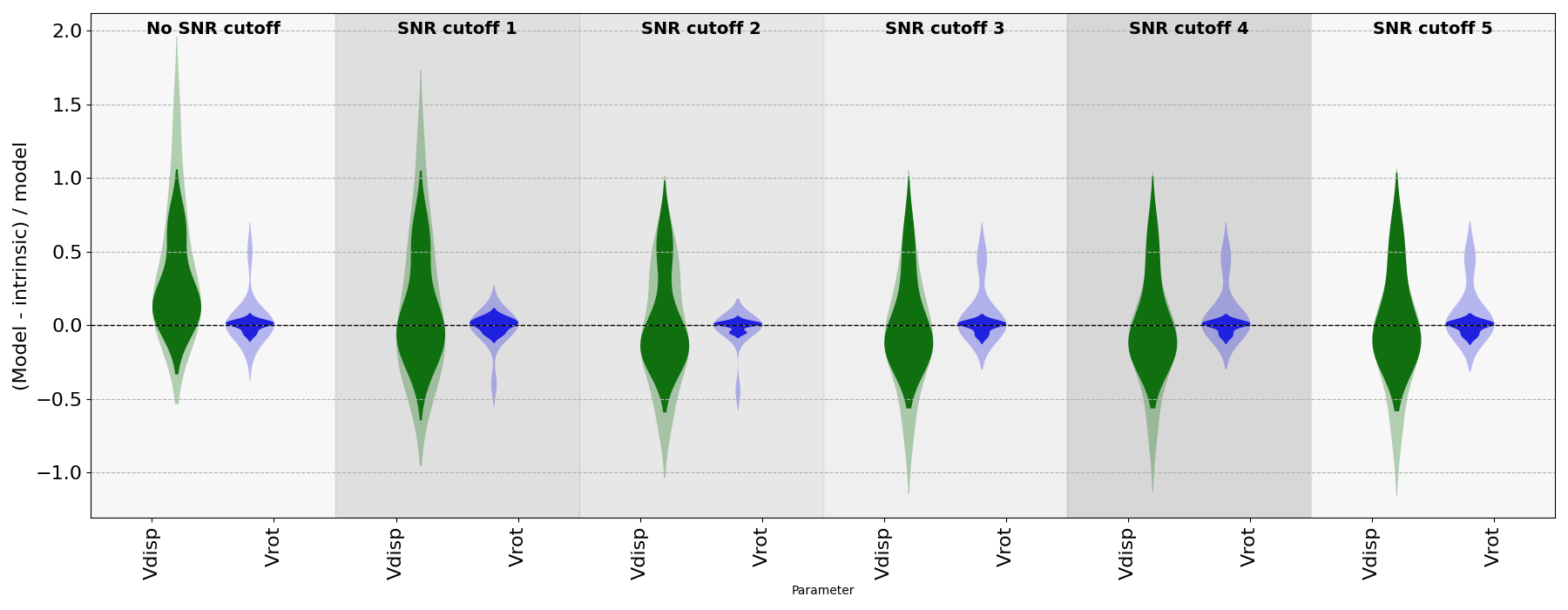}
    \caption{The impact of change of SNR cut-off value in the masking process in  \barolo. This plot shows the velocity dispersion and rotation velocity value offset from the intrinsic at $2.2 r_{\rm d}$ for the high data quality \barolo\, fits split per SNR cut-off value. The lack of variations present here remains independent of data quality and method of parameter extraction. The transparent shading shows input systems independent of rotation or dispersion domination, while the solid colours display the rotation-dominated systems only. 
    }
    \label{fig:violin_barolo_SNRvariation}
\end{figure*}


\bsp	
\label{lastpage}
\end{document}